\documentclass[twocolumn,english]{revtex4-2}
\usepackage{lmodern}

\usepackage[T1]{fontenc}
\usepackage[latin9]{inputenc}
\usepackage{babel}
\usepackage{amsmath}
\usepackage{amssymb}
\usepackage{graphicx}
\usepackage[unicode=true,pdfusetitle,
 bookmarks=true,bookmarksnumbered=false,bookmarksopen=false,
 breaklinks=false,pdfborder={0 0 1},backref=false,colorlinks=false]
 {hyperref}
\begin{document}
\title{Calculation of elastic constants of embedded-atom-model potentials
in the NVT ensamble}
\author{Menahem Krief}
\email{menahem.krief@mail.huji.ac.il}

\author{Yinon Ashkenazy}
\address{Racah Institute of Physics, The Hebrew University, 9190401 Jerusalem,
Israel}
\begin{abstract}
A method for the calculation of elastic constants in the NVT ensamble, using molecular dynamics (MD) simulation with a realistic many-body embedded-atom-model (EAM) potential, is studied in detail.
It is shown that in such NVT MD simulations, the evaluation
of elastic constants is robust and accurate, as it gives the elastic
tensor in a single simulation which converges using a small number of time steps and particles. These results highlight the
applicability of this method in: (i) the calculation
of local elastic constants of non-homogeneous  crystalline materials and (ii)
in the calibration of interatomic potentials, as a fast and accurate alternative to the common method of explicit deformation, which requires a set of consistent simulations at different conditions.
The method is demonstrated for the calculation of the elastic constants of copper in the
temperature range of 0-1000K, and results agree with the target values used for the potential calibration. The various
contributions to the values of the elastic constants, namely, the
Born, stress fluctuation and ideal gas terms, are studied as a function
of temperature. 
\end{abstract}
\maketitle

\section{Introduction}

The calculation of thermo-elastic properties of materials using computer simulations, plays a key role in understanding the
response of materials to deformation under varying conditions. In a wide range of applications, material structure leads to local variation in the elastic response functions, due to grain boundaries and hetero-phase interfaces. It is  this local variation of elastic properties that allow tailoring of the macroscopic effective material properties. The development of a model for the relation between composite material specific local property and an effective average response function depends on understanding how local properties contribute and interact in order to generate an observable average response. Furthermore, local elastic properties are not accessible experimentally for a wide range of systems, and so reliable numerical evaluation of these may play a key role in the development of effective models for composite materials.

While it is well established that various thermal and mechanical properties can be evaluated for various atomistic structures by atomistic simulations, it is also well established that such evaluation requires addressing non pairwise terms.  The widely used
embedded atom model (EAM)\cite{daw1984embedded},
is a fast, simple and accurate method, which allows a correct
description of the microscopic interactions in crystalline materials.
Molecular dynamics calculations of adiabatic elastic constants (which
are performed in the microcanonical NVE ensamble), for metallic
elements, using EAM model potentials, were performed in the past  \cite{wolf1992temperature,ccaugin1999thermal,chantasiriwan1996higher},
based on the widely used formulation of Ray et al. \cite{ray1985molecular}.

In this manuscript we present and analyze the feasibility, robustness
and accuracy of the calculation of elastic constants of metals under
constant temperature and volume, that is, in the canonical NVT ensamble.
Using the stress-stress formulation \cite{squire1969isothermal,lutsko1989generalized,clavier2017computation,lips2018stress,chantasiriwan1996higher},
all the components of the elastic constant tensor are obtained in
a single molecular dynamics simulation, as opposed to the common explicit
deformation method \cite{rassoulinejad2016evaluation,clavier2017computation,griebel2004molecular, quesnel1993elastic, manevitch2004elastic, vashishta2007interaction, pei2010mechanical, desai2019lammps}, which requires several simulations under different
deformations.
In addition, the method allows the evaluation of elastic properties in localized regions within larger non-homogeneous simulation box, unlike first principle methods which allow efficient temperature dependent calculations, but only for a uniform system \cite{huang2006thermoelastic,keuter2019qualitative}.
Finally, it is demonstrated that NVT calculations of elastic constants converge more rapidly in comparison
to the NPT strain-strain fluctuations formulation \cite{parrinello1982strain,gusev1996fluctuation,clavier2017computation,lips2018stress,shinoda2004rapid}, using 
symplectic numerical integrators \cite{martyna1994constant,martyna1996explicit,tuckerman2001non,tuckerman2006liouville,tuckerman2010statistical,allen2017computer} and 
Nose-Hoover thermostat chain \cite{martyna1992nose,nose1984unified,hoover1985canonical}.

We performed calculations of the isothermal elastic
constants $C_{11},C_{12},C_{44}$ of Copper 
for a widely used
realistic tabulated EAM potential by Mishin et al. \cite{mishin2001structural}. This potential successfully reproduces energy and stability of several nonequilibrium configurations as well as transformation
paths between different structures.
The calculations were compared with experimental results in the temperature
range 0-1000K, and a good agreement is achieved.
The manuscript is structured as follows: We start with a short review of formalism used from thermoelasticity, then we describe in detail how elastic constants can be calculated using a single MD simulation in the NVT ensamble in Section 3, and conclude with Section 4 in which we employ the described method to calculate elastic constants in Copper for a commonly used EAM potential.

\section{Thermoelasticity\label{sec:Thermoelasticity}}

We start with  a brief review of the definitions and notations of thermo-elasticity, that will be used throughout
the manuscript. Given a deformation from a reference configuration
$\boldsymbol{R}$ to a configuration $\boldsymbol{r}=\boldsymbol{r}\left(\boldsymbol{R}\right)$,
the Lagrangian strain tensor is defined by:
\begin{equation}
\eta_{\alpha\beta}=\frac{1}{2}\left(\frac{\partial r_{\gamma}}{\partial R_{\alpha}}\frac{\partial r_{\gamma}}{\partial R_{\beta}}-\delta_{\alpha\beta}\right),\label{eq:strain_def}
\end{equation}
where Greek indices such as $\alpha,\beta,\gamma$ represent Cartesian
components, for which we employ 
Einstein summation convention. In matrix notation, the strain tensor
is related to the Jacobian (metric) tensor:
\begin{equation}
J_{\alpha\beta}=\frac{\partial r_{\alpha}}{\partial R_{\beta}},
\end{equation}
 by:
\begin{equation}
\boldsymbol{\eta}=\frac{1}{2}\left(\boldsymbol{J}^{T}\boldsymbol{J}-\boldsymbol{I}\right),\label{eq:eta_J}
\end{equation}
where $\boldsymbol{I}$ is the identity matrix. 
The difference in dimensions due to deformation can thus be written using the strain
tensor \ref{eq:strain_def}  via:

\begin{equation}
\delta r^{2}-\delta R^{2}=2\eta_{\alpha\beta}\delta R_{\alpha}\delta R_{\beta}.\label{eq:strain_length}
\end{equation}
And the mechanical work due to an infinitesimal deformation with Lagrangian
strain $d\eta_{\alpha\beta}$ is given by \cite{thurston1964wave,wallace1967thermoelasticity,brugger1964thermodynamic,wallace1972thermodynamics,ray1984statistical,wojciechowski1987constant}:
\begin{equation}
\delta W/V_{0}=-\tau_{\alpha\beta}d\eta_{\alpha\beta},
\end{equation}
where $V_{0}$ is the undeformed volume and the thermodynamic tension
tensor, also known as the second Piola-Kirchhoff stress tensor \cite{thurston1964wave}, is
defined by:
\begin{equation}
\boldsymbol{\tau}=\det\left(\boldsymbol{J}\right)\boldsymbol{J}^{-1}\boldsymbol{\sigma}\boldsymbol{J}^{-T},
\end{equation}
where $\boldsymbol{\sigma}$ is the Cauchy stress tensor, and the
'$-T$' superscript denotes matrix transposition and inversion. We
note that $\boldsymbol{\tau}=\boldsymbol{\sigma}$ for a zero applied
strain ($\boldsymbol{J=}\boldsymbol{I}$, $\boldsymbol{\eta=0}$).
If the deformation process is 
reversible (an assumption which excludes plasticity),
then the first law of thermodynamics for the internal energy $\mathcal{E}$
is written as:

\begin{equation}
d\mathcal{E}=Td\mathcal{S}+V_{0}\tau_{\alpha\beta}d\eta_{\alpha\beta},
\end{equation}
where $T$ denotes the temperature and $\mathcal{S}$ the entropy.
And changes in the free energy
\begin{equation}
d\mathcal{F}=\mathcal{S}dT+V_{0}\tau_{\alpha\beta}d\eta_{\alpha\beta}.\label{eq:free_energy_change}
\end{equation}
The thermodynamic tension tensor is the thermodynamic conjugate of
the Lagrangian strain tensor:

\begin{equation}
\tau_{\alpha\beta}=\frac{1}{V_{0}}D_{\alpha\beta}\mathcal{F},\label{eq:tau_tot_F}
\end{equation}
where we used the symmetrical partial derivative operator:
\begin{equation}
D_{\alpha\beta}=\frac{1}{2}\left(\frac{\partial}{\partial\eta_{\alpha\beta}}+\frac{\partial}{\partial\eta_{\beta\alpha}}\right),
\end{equation}
due to symmetry of the strain tensor $d\eta_{\alpha\beta}=d\eta_{\beta\alpha}$
in eq. \ref{eq:free_energy_change}. The Taylor expansion of the elastic
free energy around the un-deformed state, $\boldsymbol{\eta=0}$, is
written as:

\begin{equation}
\mathcal{F}\left(\boldsymbol{\eta}\right)/V_{0}=\mathcal{F}\left(\boldsymbol{0}\right)/V_{0}+\sigma_{\alpha\beta}\eta_{\alpha\beta}+\frac{1}{2}C_{\alpha\beta\gamma\delta}\eta_{\alpha\beta}\eta_{\gamma\delta}+...
\end{equation}
where:

\begin{equation}
\sigma_{\alpha\beta}=\tau_{\alpha\beta}\left(\boldsymbol{\eta=0}\right)=\frac{1}{V_{0}}D_{\alpha\beta}\mathcal{F}\bigg|_{\boldsymbol{\eta}=\boldsymbol{0}},\label{eq:sigma_cauchy}
\end{equation}
and the second order isothermal elastic constants are defined by:
\begin{equation}
C_{\alpha\beta\gamma\delta}=\frac{1}{V_{0}}D_{\alpha\beta}D_{\gamma\delta}\mathcal{F}\bigg|_{\boldsymbol{\eta}=\boldsymbol{0}},\label{eq:elastic_def}
\end{equation}
where the derivatives in equations \ref{eq:sigma_cauchy}-\ref{eq:elastic_def}
are taken at constant $T$ and evaluated at zero
strain $\boldsymbol{\eta}=\boldsymbol{0}$.

\section{molecular dynamics and Elastic constants\label{sec:molecular-dynamics-and}}
In this section we briefly outline how the entire isothermal elasticity tensor can be obtained in a single molecular dynamics simulation using the stress-stress fluctuation method \cite{ray1984statistical,ray1985molecular,ray1988elastic,wolf1992temperature,clavier2017computation,lips2018stress}. This method is favorable over the widely used explicit deformation method \cite{clavier2017computation,griebel2004molecular, quesnel1993elastic, manevitch2004elastic, vashishta2007interaction, pei2010mechanical, desai2019lammps}, which requires multiple simulations under various deformations, in order to obtain different components of the elasticity tensor.

Starting from the Hamiltonian of a system of $N$ particles is written as:

\begin{align}
\mathcal{H}\left(\boldsymbol{r}^{N},\boldsymbol{p}^{N}\right) & =\sum_{i}\frac{p_{i}^{2}}{2m_{i}}+\mathcal{V}\left(\boldsymbol{r}_{1},...,\boldsymbol{r}_{N}\right),\label{eq:hamiltonian}
\end{align}
where $m_{i}$,$\boldsymbol{r}^{N}$ and $\boldsymbol{p}^{N}$ denotes,
respectively, the masses, position and momentum vectors of the $N$
particles. The embedded-atom-model (EAM) potential \cite{daw1984embedded},
is defined by a pair potential $v=v\left(r\right)$, an embedding
function $F=F\left(\rho\right)$ and a local electron density function $\rho=\rho\left(r\right)$,
so that the the potential energy takes the form:

\begin{equation}
\mathcal{V}\left(\boldsymbol{r}_{1},...,\boldsymbol{r}_{N}\right)=\sum_{i}F\left(\rho_{i}\right)+\sum_{i<j}v\left(r_{ij}\right),\label{eq:EAM}
\end{equation}
where $\boldsymbol{r}_{ij}=\boldsymbol{r}_{i}-\boldsymbol{r}_{j}$
is the interatomic displacement vector and the electron density function of particle $i$ is given by
$\rho_{i}=\sum_{j\neq i}\rho\left(r_{ij}\right).$
For an EAM potential of the form \ref{eq:EAM}, the force on a particle $i$ can be written
as:
\begin{equation}
\boldsymbol{F}_{i}=\sum_{j\neq i}F_{ij}\frac{\boldsymbol{r}_{ij}}{r_{ij}},\label{eq:Fi}
\end{equation}
where:
\begin{equation}
F_{ij}=-\left(v'\left(r_{ij}\right)+\left[F'\left(\rho_{i}\right)+F'\left(\rho_{j}\right)\right]\rho'\left(r_{ij}\right)\right).\label{eq:Fij_eam}
\end{equation}


The instantaneous kinetic temperature is given by:

\begin{equation}
\frac{1}{2}k_{B}T_{K}=\frac{1}{g}\sum_{i}\frac{p_{i}^{2}}{2m_{i}},\label{eq:temp_kin}
\end{equation}
where $k_{B}$ is Boltzmann's constant and $g=3\left(N-1\right)$
is the number of degrees of freedom. The instantaneous pressure tensor
$P_{\alpha\beta}$ is obtained from the virial theorem:

\begin{align*}
P_{\alpha\beta}V & =\sum_{i}\left(\frac{p_{i,\alpha}p_{i,\beta}}{m_{i}}+r_{i,\alpha}F_{i,\beta}\right).
\end{align*}
Where $V$ is the system volume. The total pressure is given by the
average of the diagonal components:
\begin{equation}
P=\frac{1}{3}\left(P_{11}+P_{22}+P_{33}\right).\label{eq:total_pressure}
\end{equation}
For an EAM potential, it follows from eq. \ref{eq:Fi} that:
\begin{align*}
\sum_{i}r_{i,\alpha}F_{i,\beta} & =\sum_{i<j}F_{ij}\frac{r_{ij,\alpha}r_{ij,\beta}}{r_{ij}}
\end{align*}
so that the pressure tensor has the form:

\begin{align}
P_{\alpha\beta}V & =\sum_{i}\frac{p_{i,\alpha}p_{i,\beta}}{m_{i}}+\sum_{i<j}F_{ij}\frac{r_{ij,\alpha}r_{ij,\beta}}{r_{ij}}\label{eq:pressure_tensor}
\end{align}
It can be shown \cite{squire1969isothermal,lutsko1989generalized,ccaugin1999thermal,van2005isothermal,barrat2006microscopic,clavier2017computation,lips2018stress}, as detailed in Appendix \ref{app:Elastic-constant},
that in the NVT ensamble, the elastic constants \ref{eq:elastic_def}
are given by:

\begin{align}
C_{\alpha\beta\gamma\delta} & =\left\langle C_{\alpha\beta\gamma\delta}^{B}\right\rangle -\frac{V}{k_{B}T}\left[\left\langle \sigma_{\alpha\beta}^{B}\sigma_{\gamma\delta}^{B}\right\rangle -\left\langle \sigma_{\alpha\beta}^{B}\right\rangle \left\langle \sigma_{\gamma\delta}^{B}\right\rangle \right]\nonumber \\
 & +\frac{Nk_{B}T}{V}\left(\delta_{\alpha\gamma}\delta_{\beta\delta}+\delta_{\alpha\delta}\delta_{\beta\gamma}\right),\label{eq:elastic_micro}
\end{align}
where $\left\langle \cdot\right\rangle $ represents ensemble average.
The last term in eq. \ref{eq:elastic_micro} is a non configurational
ideal gas contribution which vanishes at zero temperature and is related
to volume derivatives with respect to the strain tensor. The first
term in eq. \ref{eq:elastic_micro}, known as the Born term, is a
configurational part which is given by a canonical average of the
zero-temperature expression for the elastic constant \cite{born1954dynamical,clavier2017computation,ray1985molecular}.
The second term in eq. \ref{eq:elastic_micro} accounts for stress
fluctuations and also vanishes at zero temperature (this term is obtained
directly from the general identity  \ref{eq:strain_deriv_A} detailed in Appendix \ref{app:Elastic-constant}). The
Born stress tensor in eq. \ref{eq:elastic_micro} is defined by the
derivative of the potential energy with respect to strain, evaluated
at a state of zero applied strain:
\begin{equation}
\sigma_{\alpha\beta}^{B}=\frac{1}{V}D_{\alpha\beta}\mathcal{V}\bigg|_{\boldsymbol{\eta}=\boldsymbol{0}}.\label{eq:sigma_born}
\end{equation}
The Cauchy stress tensor is given by:
\begin{equation}
\sigma_{\alpha\beta}=\left\langle \sigma_{\alpha\beta}^{B}\right\rangle -\frac{k_{B}TN}{V}\delta_{\alpha\beta}\label{eq:sigma_cauchy_id}
\end{equation}
Similarly, the Born elastic constant term is given by:

\begin{equation}
C_{\alpha\beta\gamma\delta}^{B}=\frac{1}{V}D_{\alpha\beta}D_{\gamma\delta}\mathcal{V}\bigg|_{\boldsymbol{\eta}=\boldsymbol{0}}.
\end{equation}
For an EAM potential of the form \ref{eq:EAM}, it can be shown that
the Born stress has the following form:

\begin{equation}
V\sigma_{\alpha\beta}^{B}=-\sum_{i<j}F_{ij}\frac{r_{ij,\alpha}r_{ij,\beta}}{r_{ij}},\label{eq:sigma_born_F}
\end{equation}
while the Born elastic constant takes the form \cite{lutsko1989generalized,wolf1992temperature,chantasiriwan1996higher,ccaugin1999thermal}:
\begin{align}
VC_{\alpha\beta\gamma\delta}^{B} & =\sum_{i<j}X_{ij}\frac{r_{ij,\alpha}r_{ij,\beta}r_{ij,\gamma}r_{ij,\delta}}{r_{ij}^{2}}\nonumber \\
 & +\sum_{i}F''\left(\rho_{i}\right)g_{i,\alpha\beta}g_{i,\gamma\delta},\label{eq:cborn_micro}
\end{align}
where:

\begin{align}
X_{ij} & =v''\left(r_{ij}\right)-\frac{1}{r_{ij}}v'\left(r_{ij}\right)\nonumber \\
 & +\left(F'\left(\rho_{i}\right)+F'\left(\rho_{j}\right)\right)\left(\rho''\left(r_{ij}\right)-\frac{1}{r_{ij}}\rho'\left(r_{ij}\right)\right),\label{eq:Xij}
\end{align}
and:
\begin{equation}
g_{i,\alpha\beta}=\sum_{k\neq i}\rho'\left(r_{ik}\right)\frac{r_{ik,\alpha}r_{ik,\beta}}{r_{ik}}.\label{eq:g_def}
\end{equation}
For completeness, a detailed derivation of equations \ref{eq:elastic_micro}-\ref{eq:g_def},
is given in Appendix \ref{app:Elastic-constant}.

\section{Results\label{sec:Results}}

\begin{figure}
\begin{centering}
\includegraphics[scale=0.5]{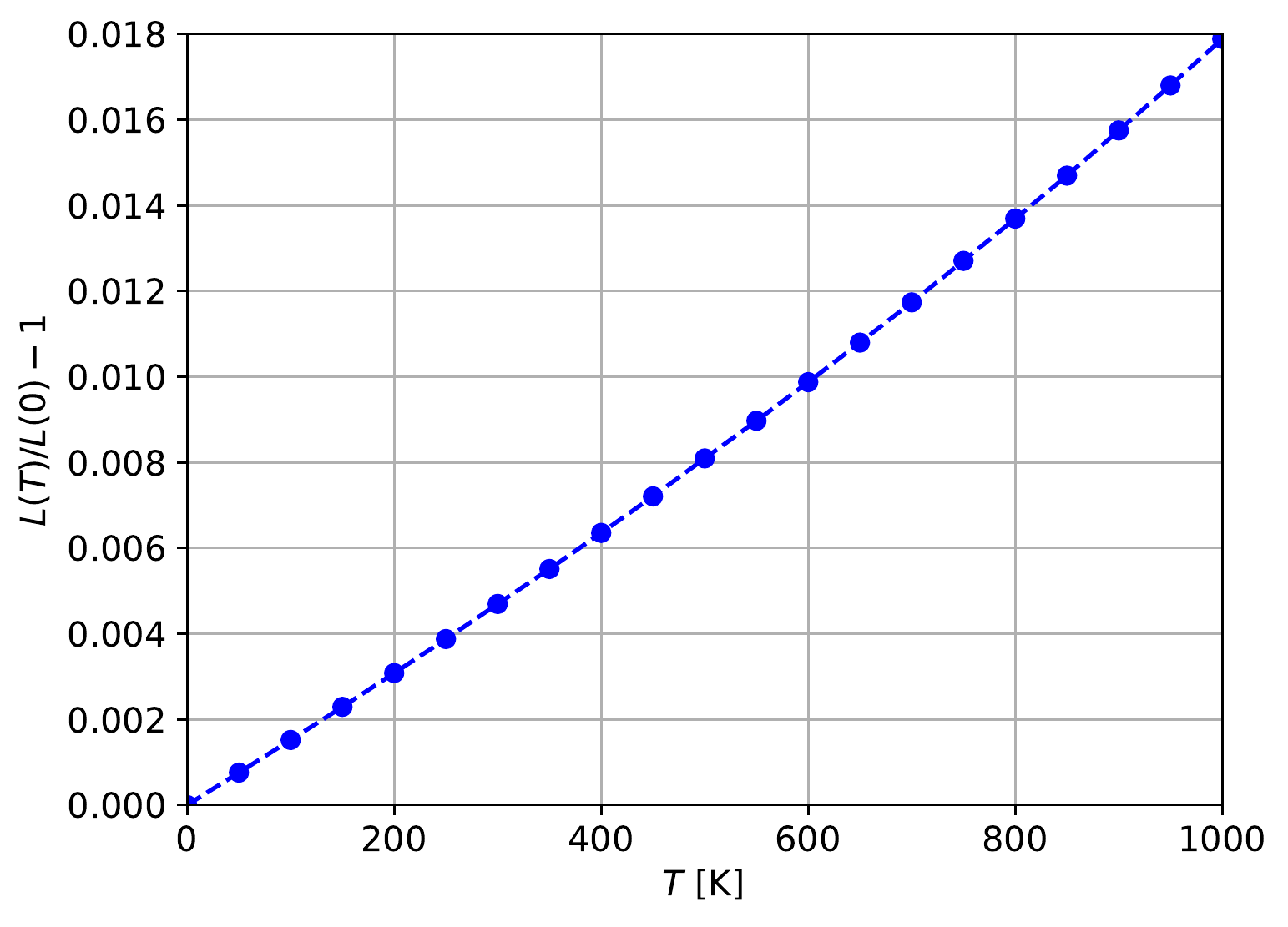}
\par\end{centering}
\caption{The thermal expansion ratio $L\left(T\right)/L\left(0\right)-1$ of
Copper, as a function of temperature, resulting from a series of NPT
molecular dynamics simulations at zero pressure in the temperature
range 0-1000K. Time dependent measures of the simulation at T=300K
are presented in detail in figures \ref{fig:npt_time}-\ref{fig:npt_p_tensor}.
\label{fig:L_T}}
\end{figure}

\begin{figure*}
\begin{centering}
\includegraphics[scale=0.38]{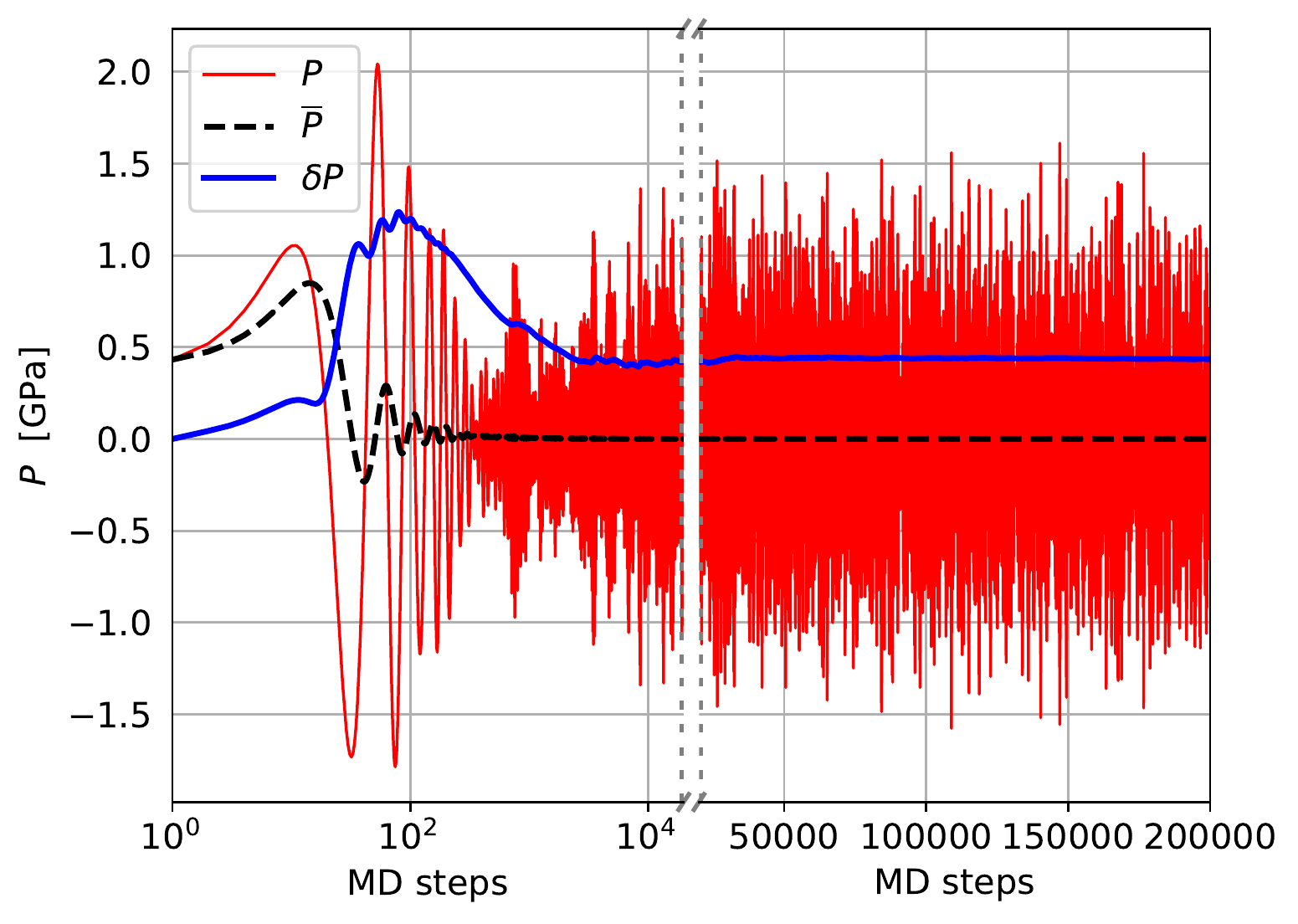}\includegraphics[scale=0.38]{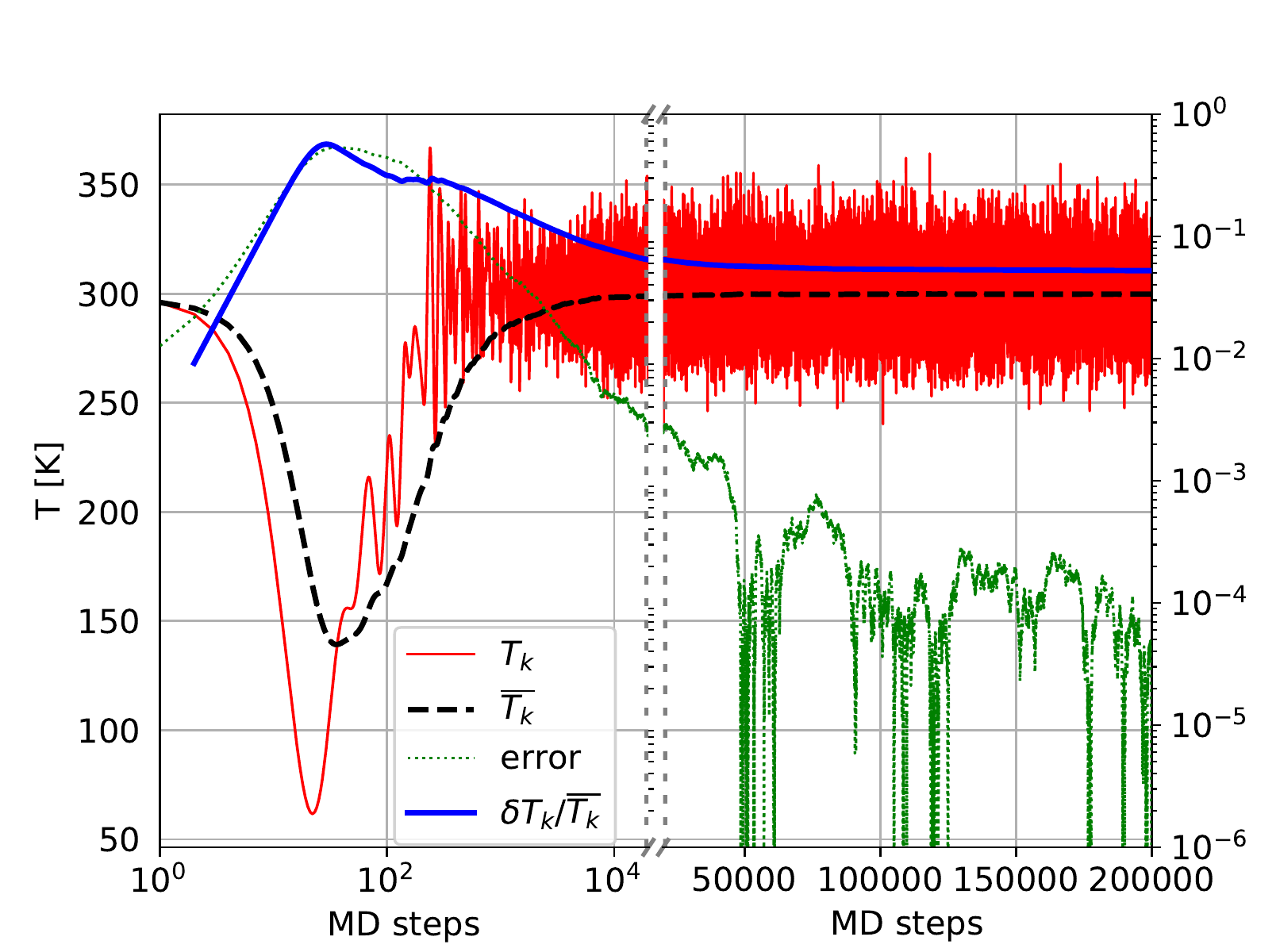}\includegraphics[scale=0.38]{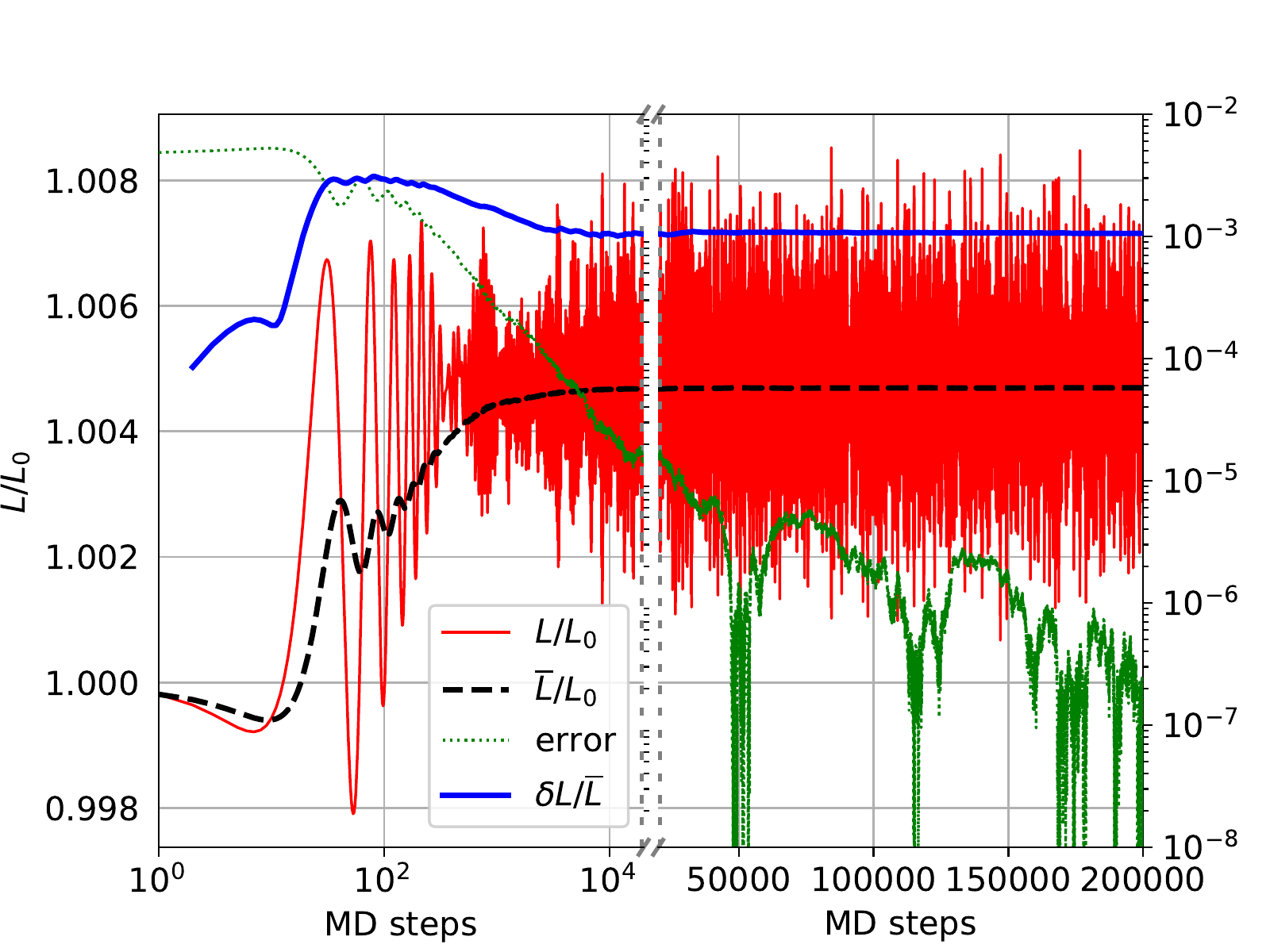}
\par\end{centering}
\begin{centering}
\includegraphics[scale=0.38]{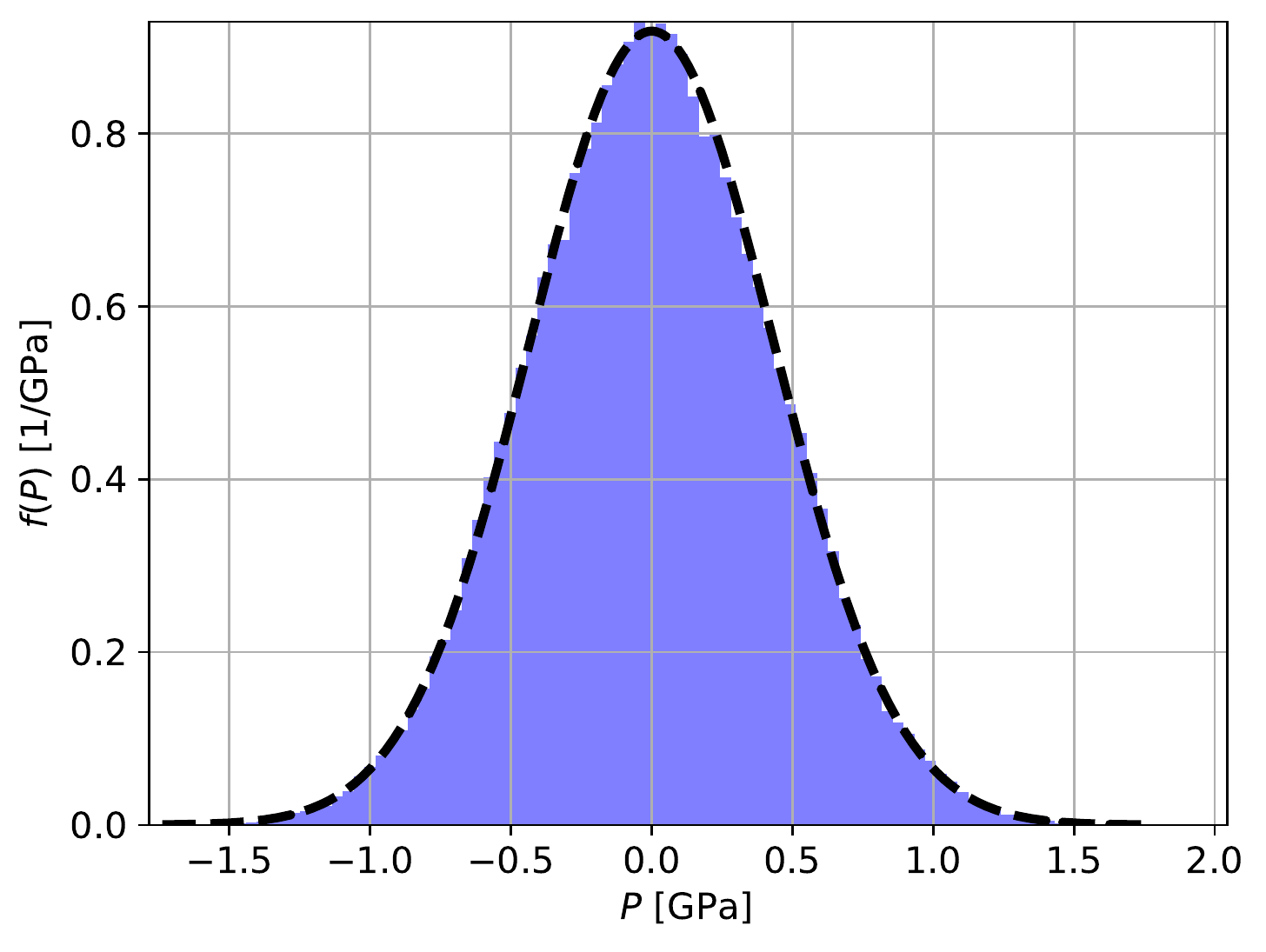}\includegraphics[scale=0.38]{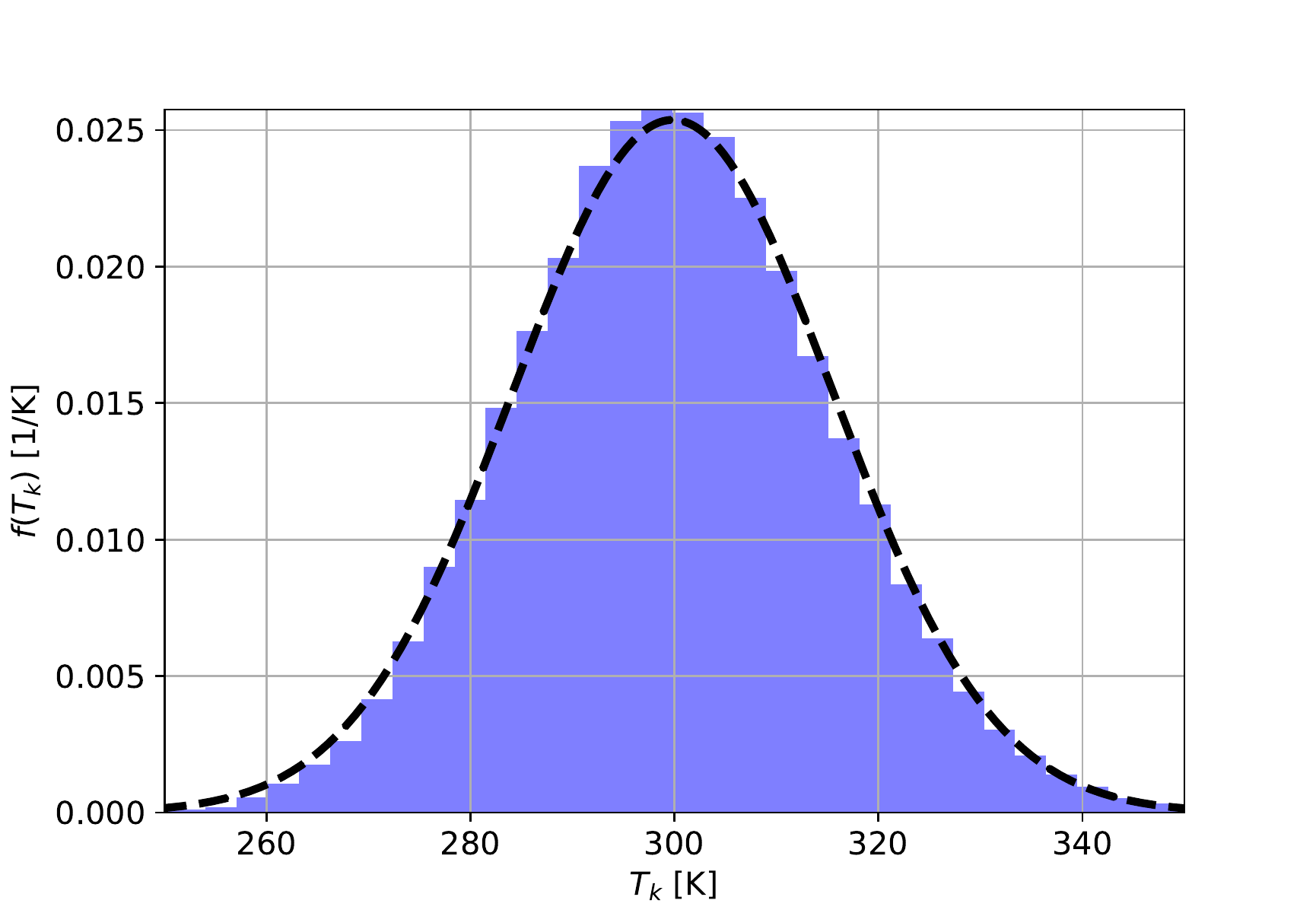}\includegraphics[scale=0.38]{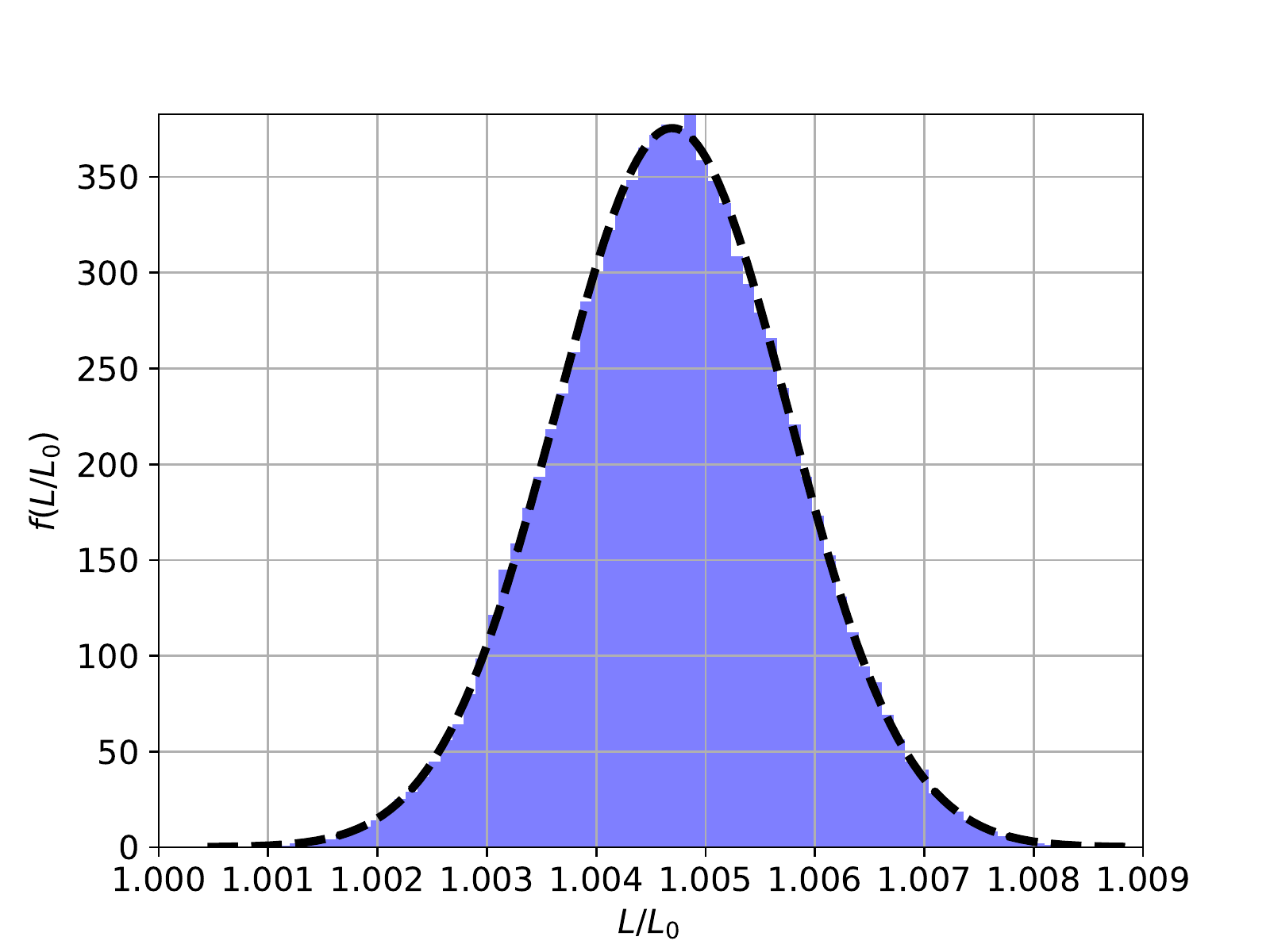}
\par\end{centering}
\caption{(Color online) Analysis of a molecular dynamics simulation of Copper in the NPT ensemble
at zero pressure and $T=300\text{K}$. The upper left figure shows
the instantaneous pressure (eq. \ref{eq:total_pressure}, red solid
line), the cumulative average pressure (black dashed line) and the
cumulative pressure standard deviation. The upper middle figure shows
the instantaneous kinetic temperature (eq. \ref{eq:temp_kin}, red
solid line, left y-axis), the cumulative average temperature (black
dashed line, left y-axis), the relative error between the current
cumulative average value to the the final average value (green dotted
line) and the ratio between the cumulative standard deviation to the
cumulative average (blue solid line, right axis). Similarly, the upper
right figure shows the results for the ratio between the simulation
box length and the initial box length ($L_{0}=3.165\mathrm{\mathring{A}}$).
The first $2\times10^{4}$ steps are plotted on a logarithmic x scale,
in order to show the initial thermalization period. The lower figures
show histograms and fitted Gaussian distributions for the values of
the instantaneous pressure (lower left), temperature (middle figure)
and box size ratio (lower right).\label{fig:npt_time}}
\end{figure*}

\begin{figure}
\begin{centering}
\includegraphics[scale=0.45]{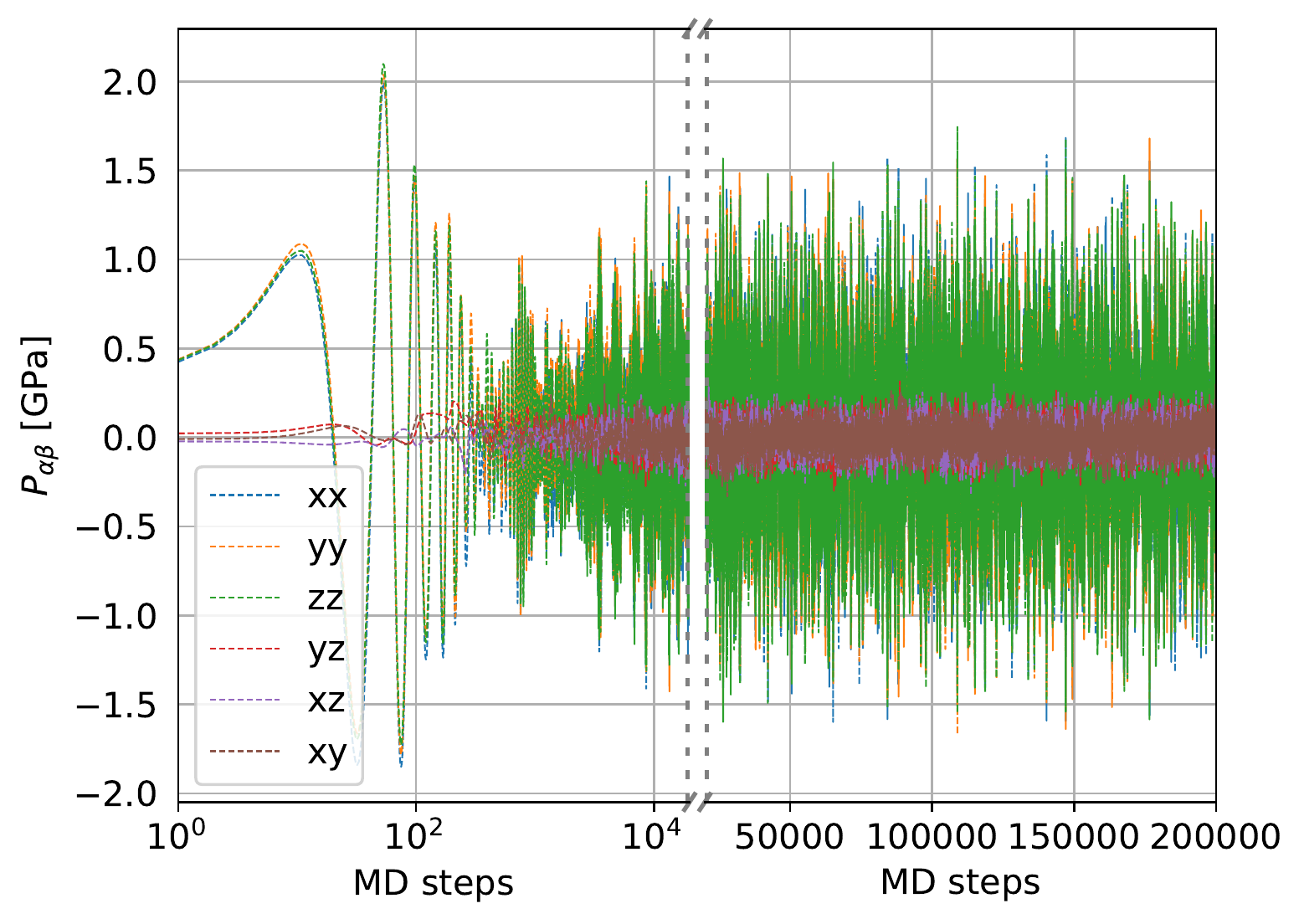}
\par\end{centering}
\caption{(Color online) Components of the instantaneous pressure tensor (eq. \ref{eq:temp_kin})
for the NPT simulation described in Fig. \ref{fig:npt_time}.\label{fig:npt_p_tensor}}
\end{figure}

\begin{figure}
\begin{centering}
\includegraphics[scale=0.45]{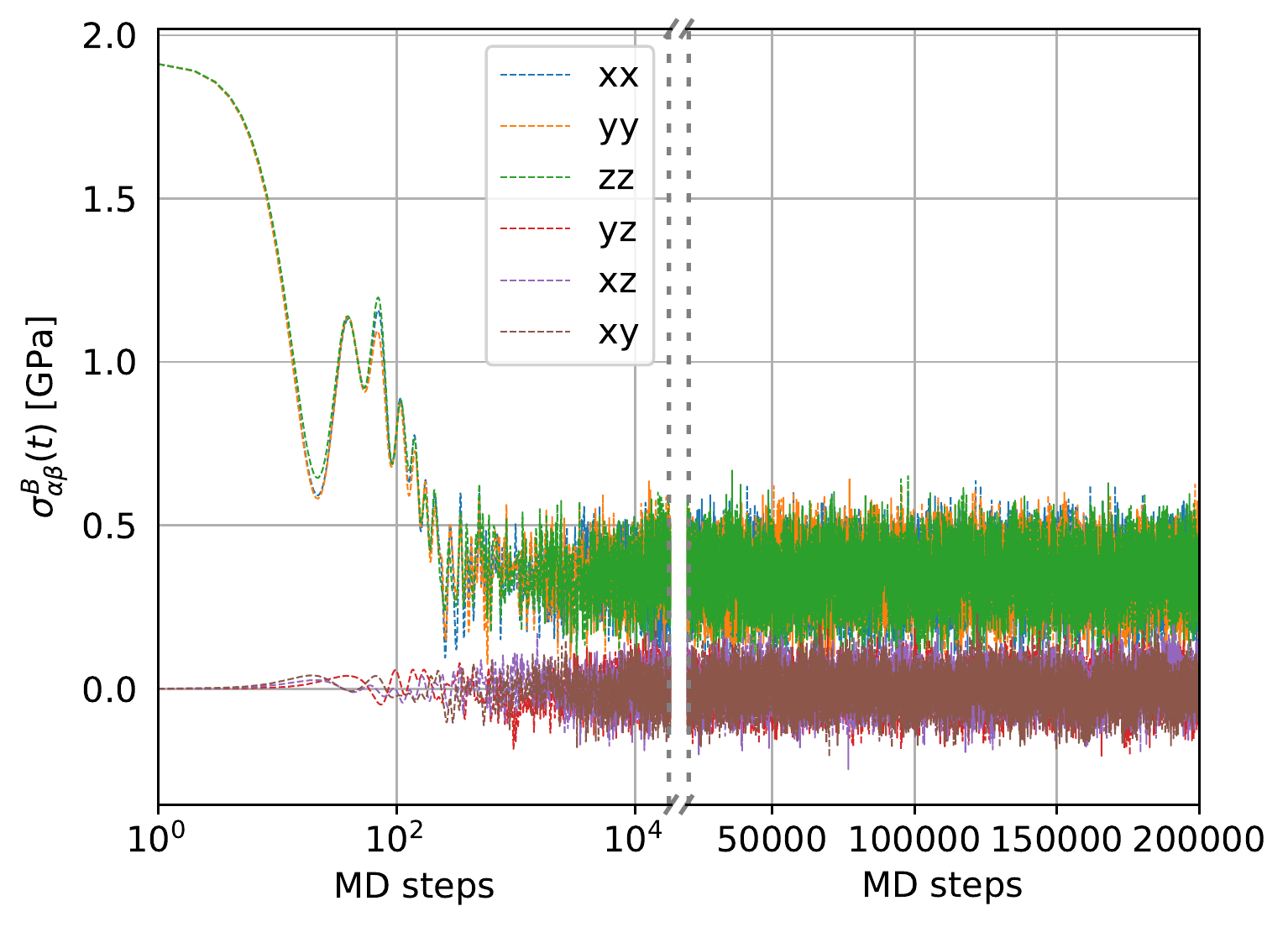}
\par\end{centering}
\begin{centering}
\includegraphics[scale=0.45]{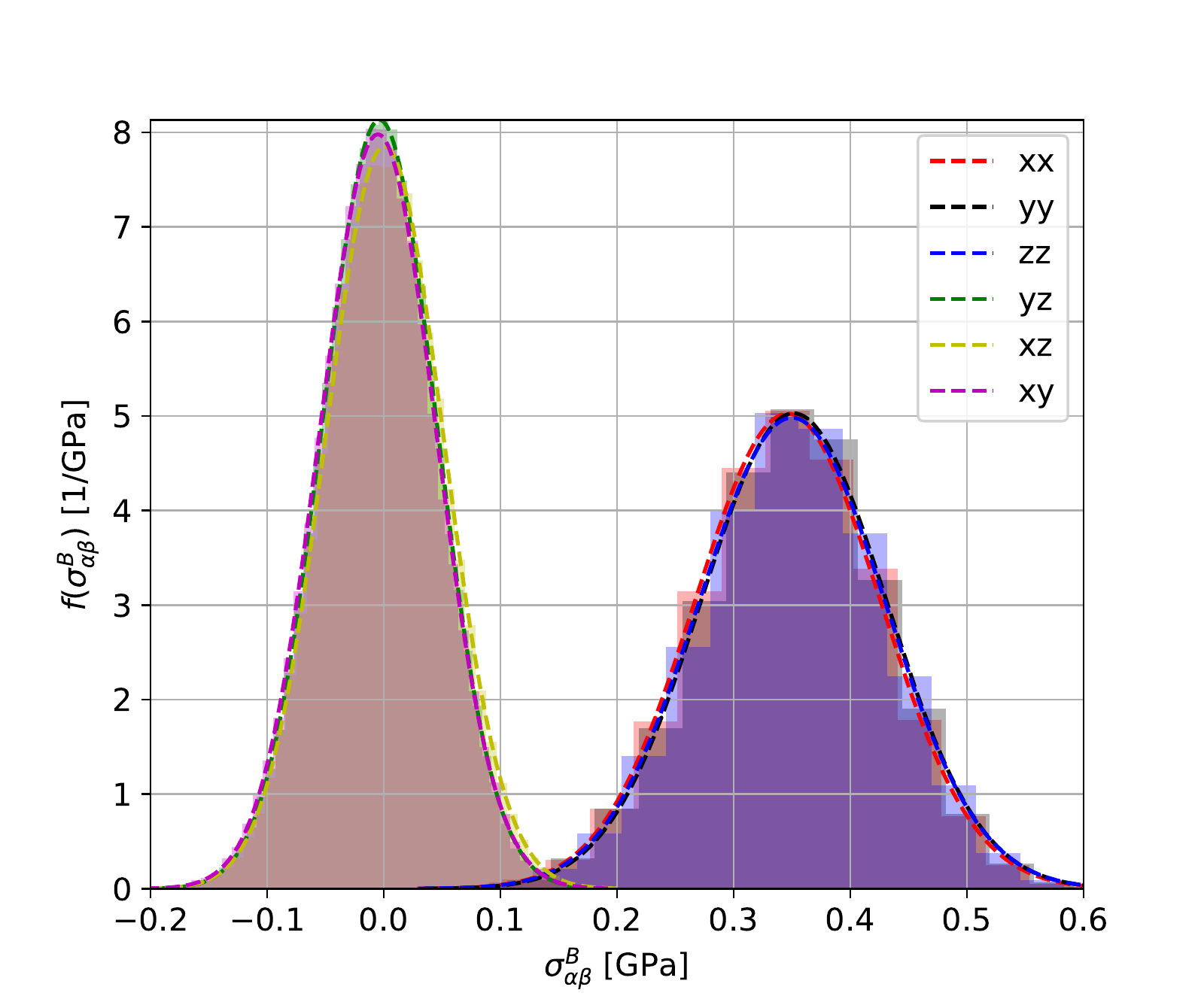}
\par\end{centering}
\caption{(Color online) Components of the instantaneous Born stress tensor (upper figure,
eq. \ref{eq:sigma_born_F}) for an NVT simulation of Copper at $T=300$K
for which the volume is chosen such that the total pressure is zero
(as obtained from Fig. \ref{fig:L_T}). The lower figure shows the
resulting histograms and fitted Gaussian distributions of the various
Born stress tensor components.\label{fig:NVT_born_stress_fluc}}
\end{figure}

\begin{figure*}
\begin{centering}
\includegraphics[scale=0.38]{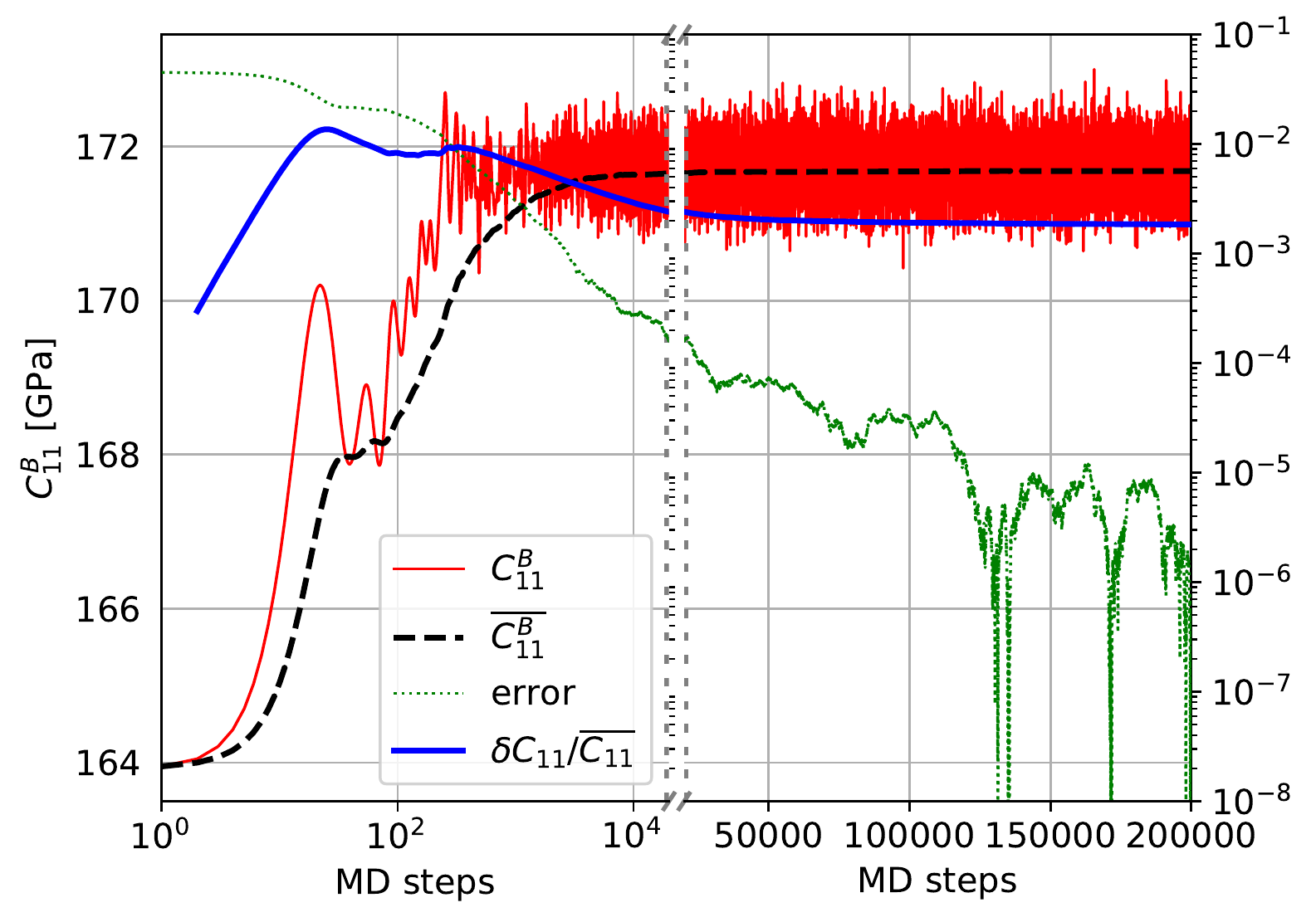}\includegraphics[scale=0.38]{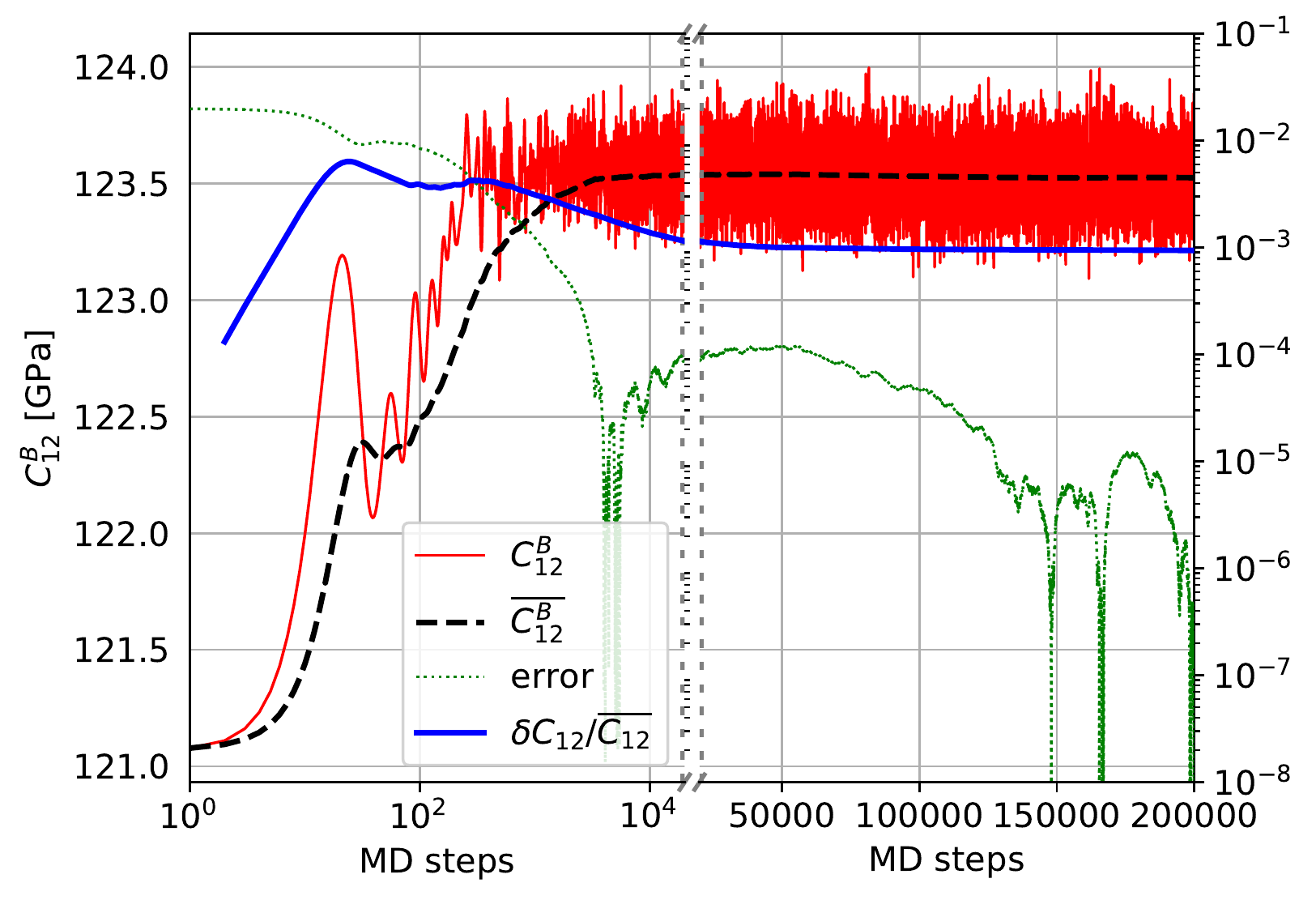}\includegraphics[scale=0.38]{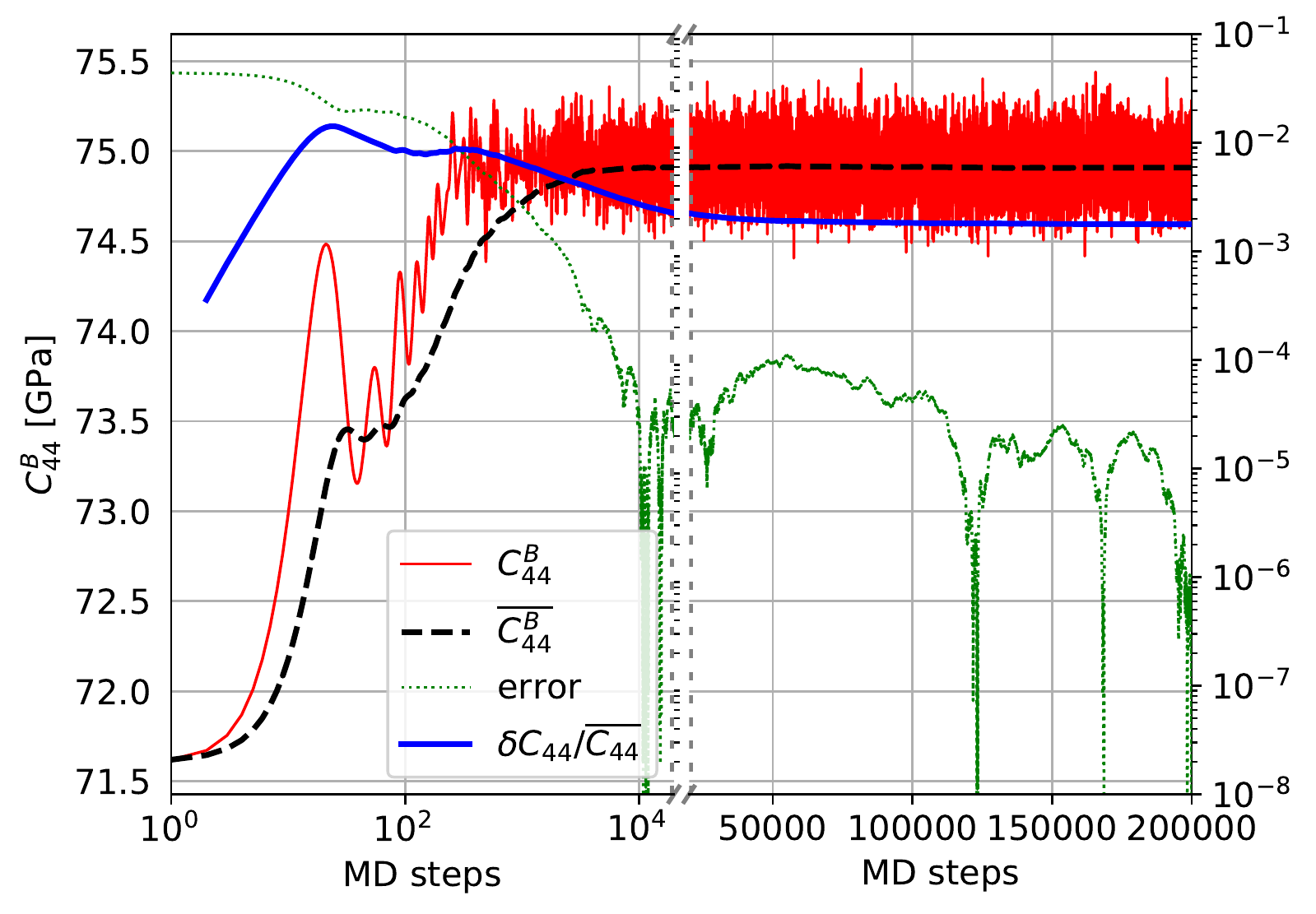}
\par\end{centering}
\begin{centering}
\includegraphics[scale=0.38]{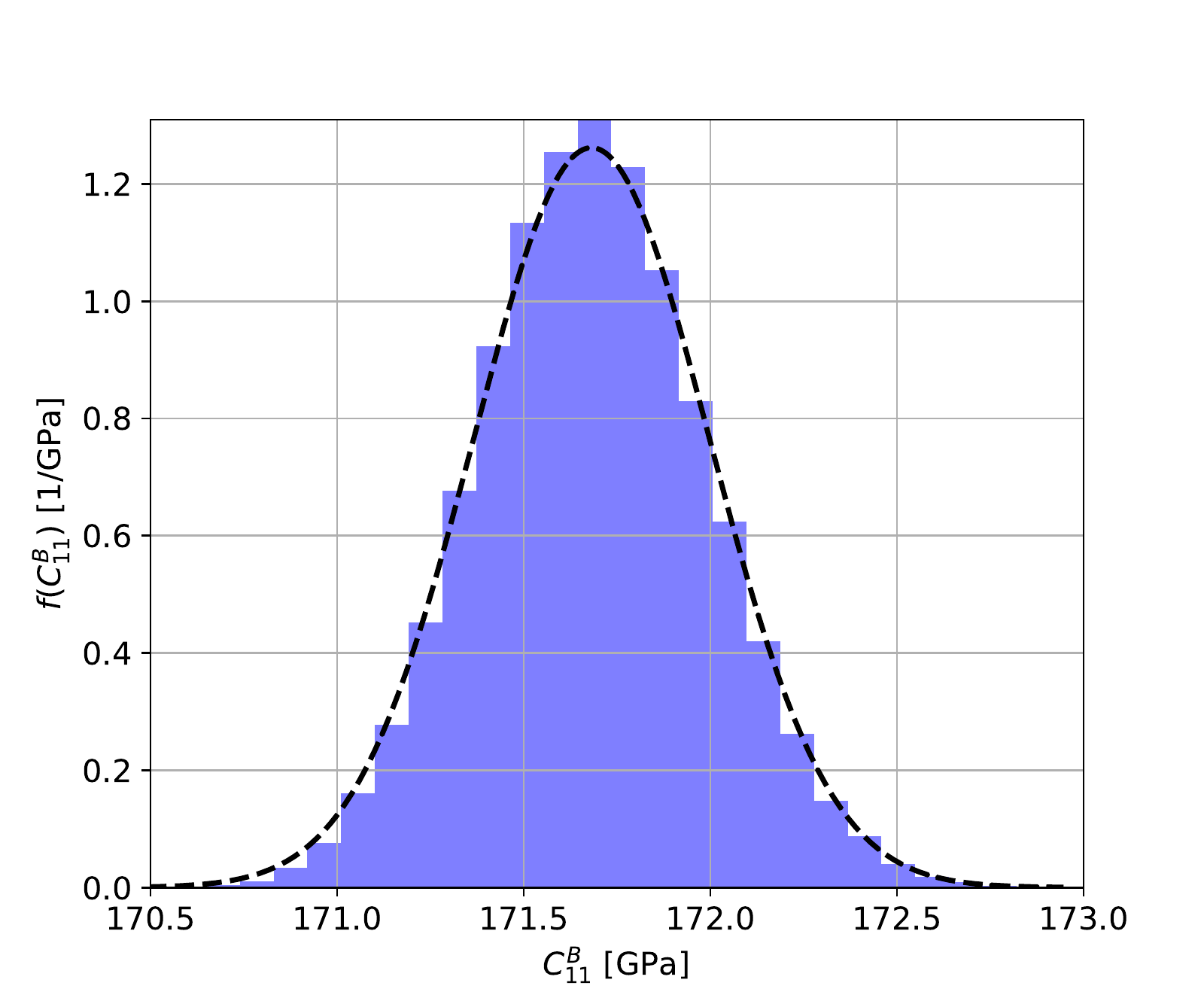}\includegraphics[scale=0.38]{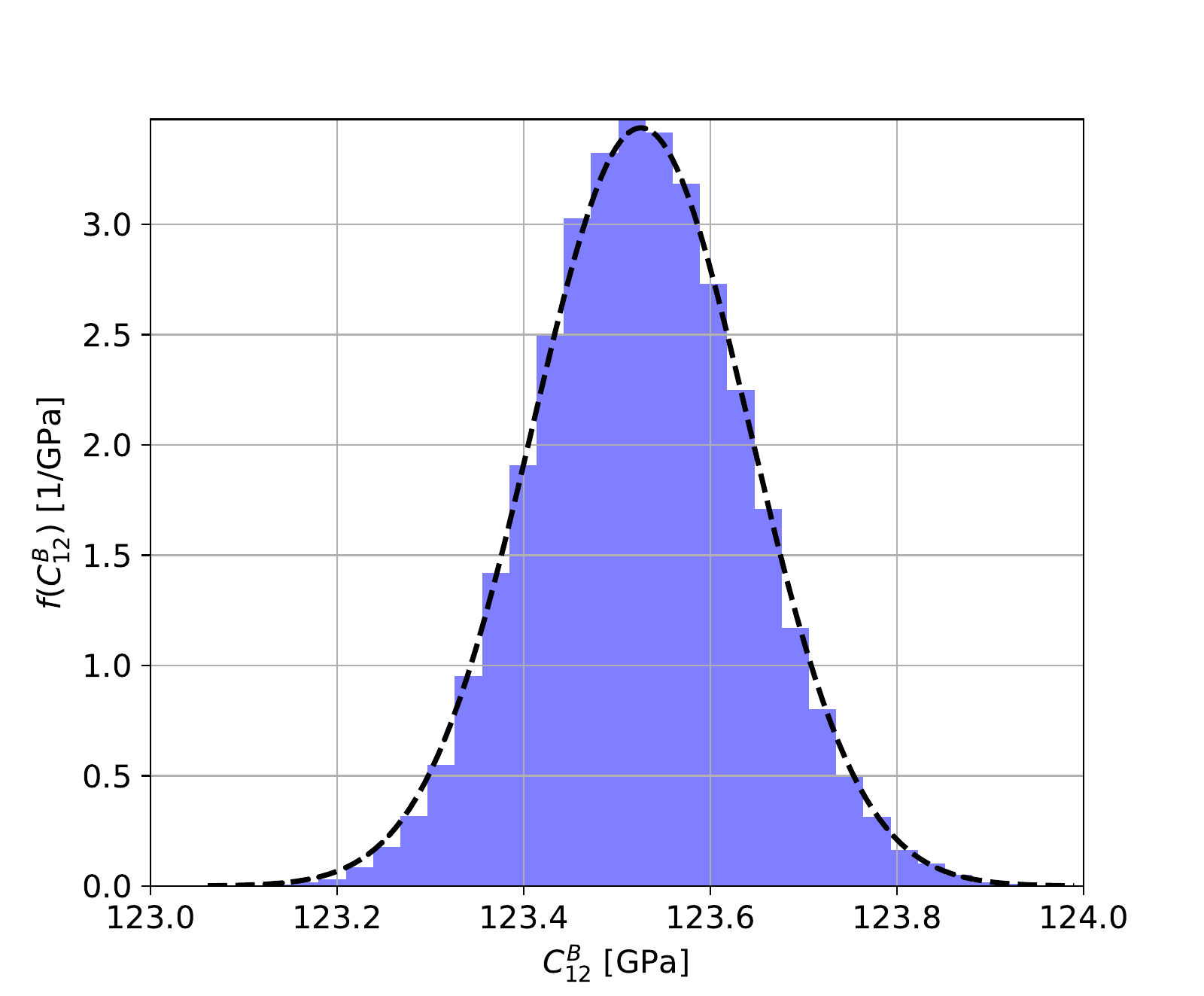}\includegraphics[scale=0.38]{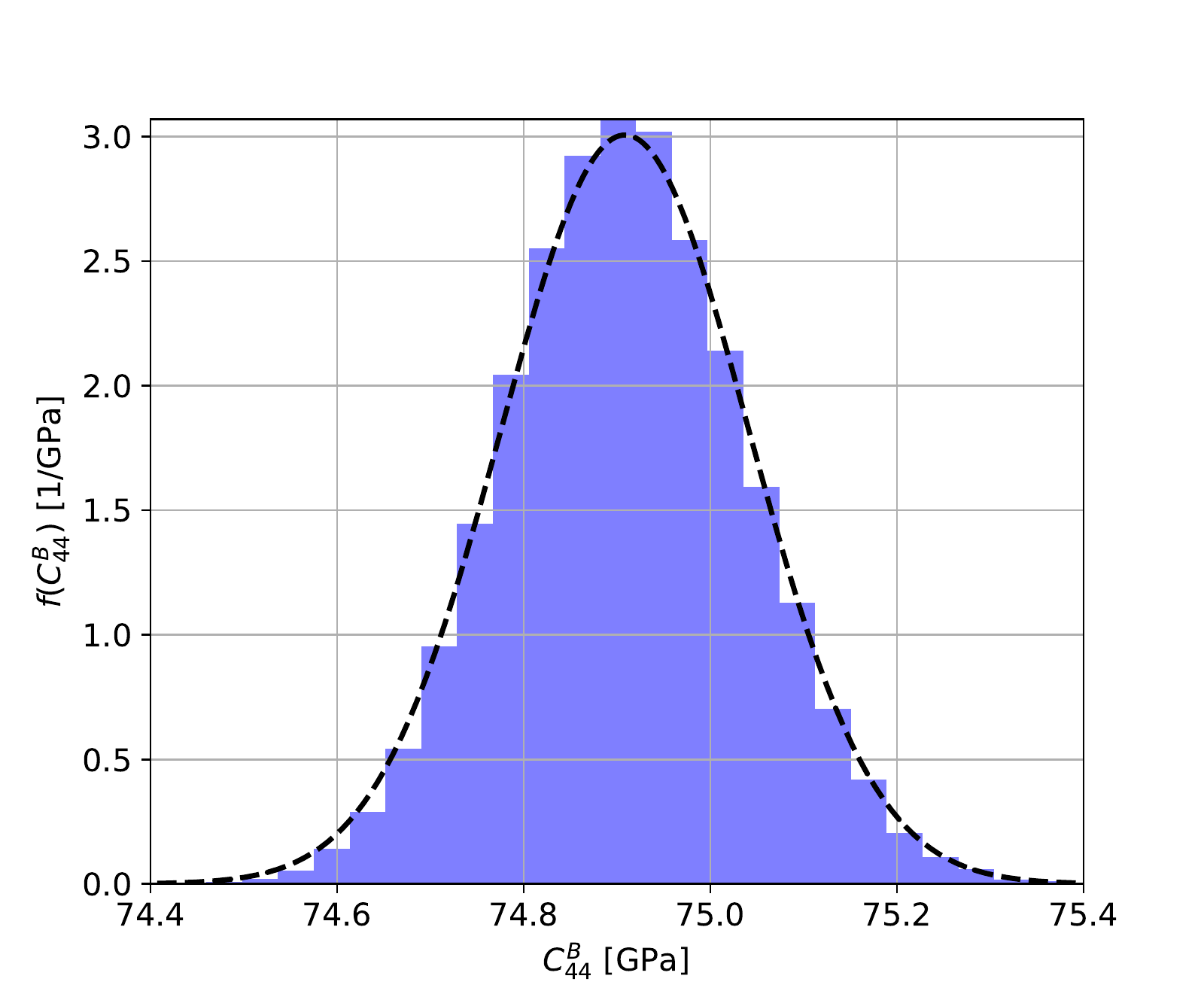}
\par\end{centering}
\caption{(Color online) Upper figures - the Born elastic constants (eq. \ref{eq:cborn_micro}):
$C_{11}^{B}$ (left upper figure), $C_{12}^{B}$ (middle upper figure)
and $C_{44}^{B}$ (right upper figure), for an NVT simulation of Copper
at $T=300$K (as described in Fig. \ref{fig:NVT_born_stress_fluc}).
The instantaneous value, cumulative average, convergence error and
ratio of the standard deviation to the mean are shown, as is detailed
in Fig. \ref{fig:npt_time}. The lower figures shows the resulting
histograms and fitted Gaussian distributions.\label{fig:NVT_C_born}}
\end{figure*}

\begin{figure}
\begin{centering}
\includegraphics[scale=0.5]{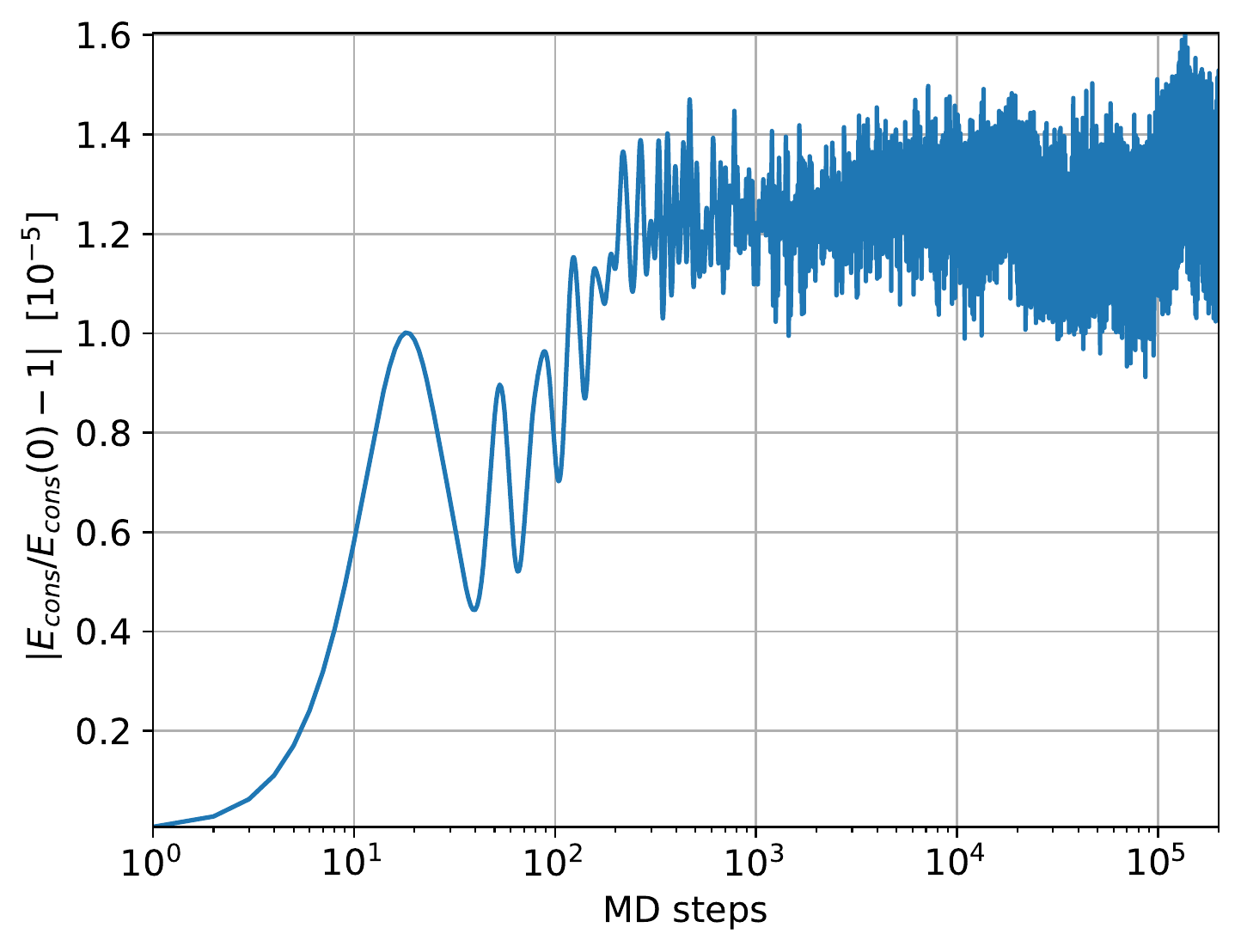}
\par\end{centering}
\begin{centering}
\includegraphics[scale=0.5]{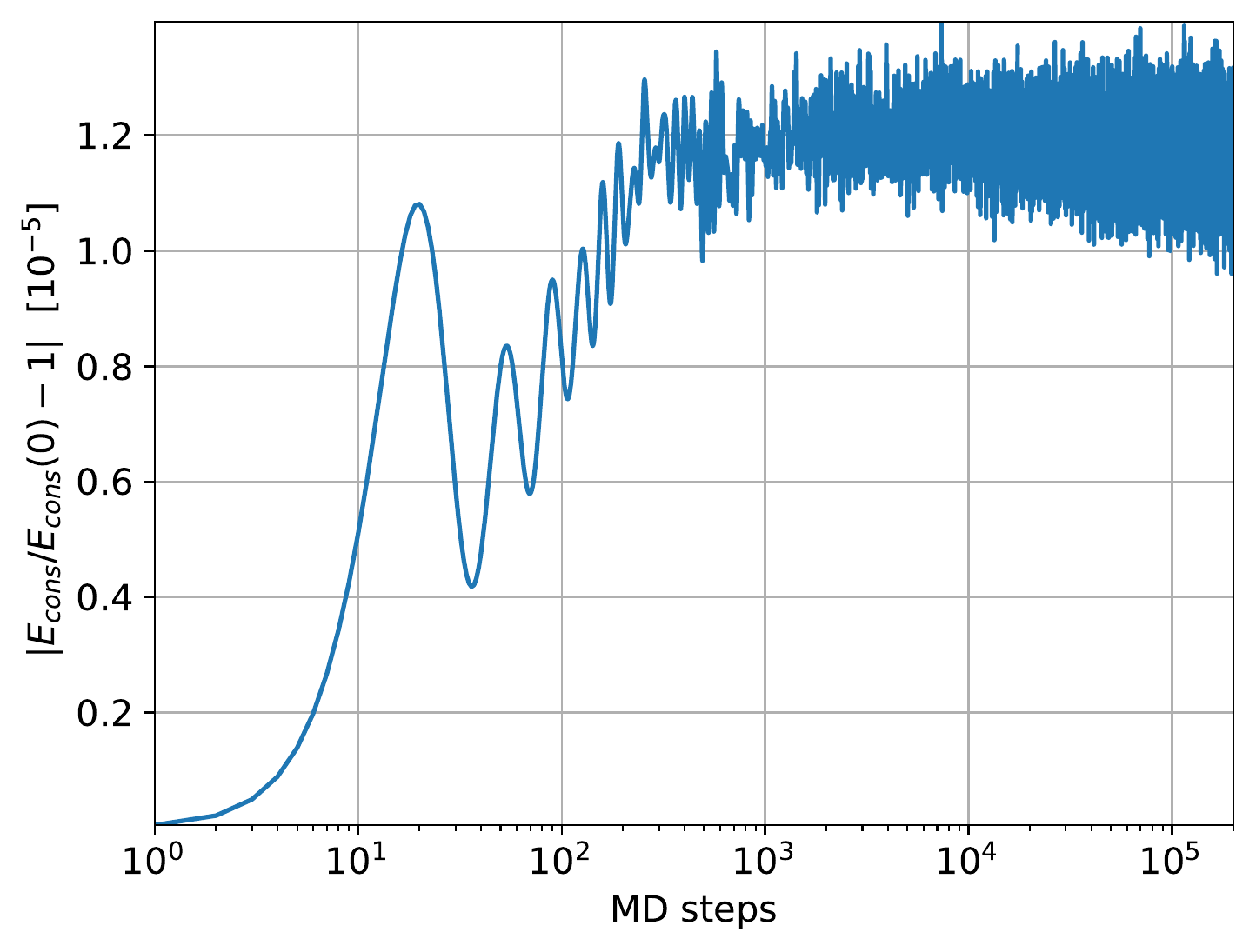}
\par\end{centering}
\caption{Relative error between the instantaneous conserved energy and the
initial value, for NPT (upper figure, eq. \ref{eq:econst_nvt}) and
NVT (upper figure, eq. \ref{eq:econst_npt}) molecular dynamics simulations
of Copper at $T=300$K and zero pressure (the simulations described
in figures \ref{fig:npt_time}-\ref{fig:npt_p_tensor} and \ref{fig:NVT_born_stress_fluc}-\ref{fig:NVT_C_born},
respectively).\label{fig:econs}}
\end{figure}

\begin{figure*}
\begin{centering}
\includegraphics[scale=0.38]{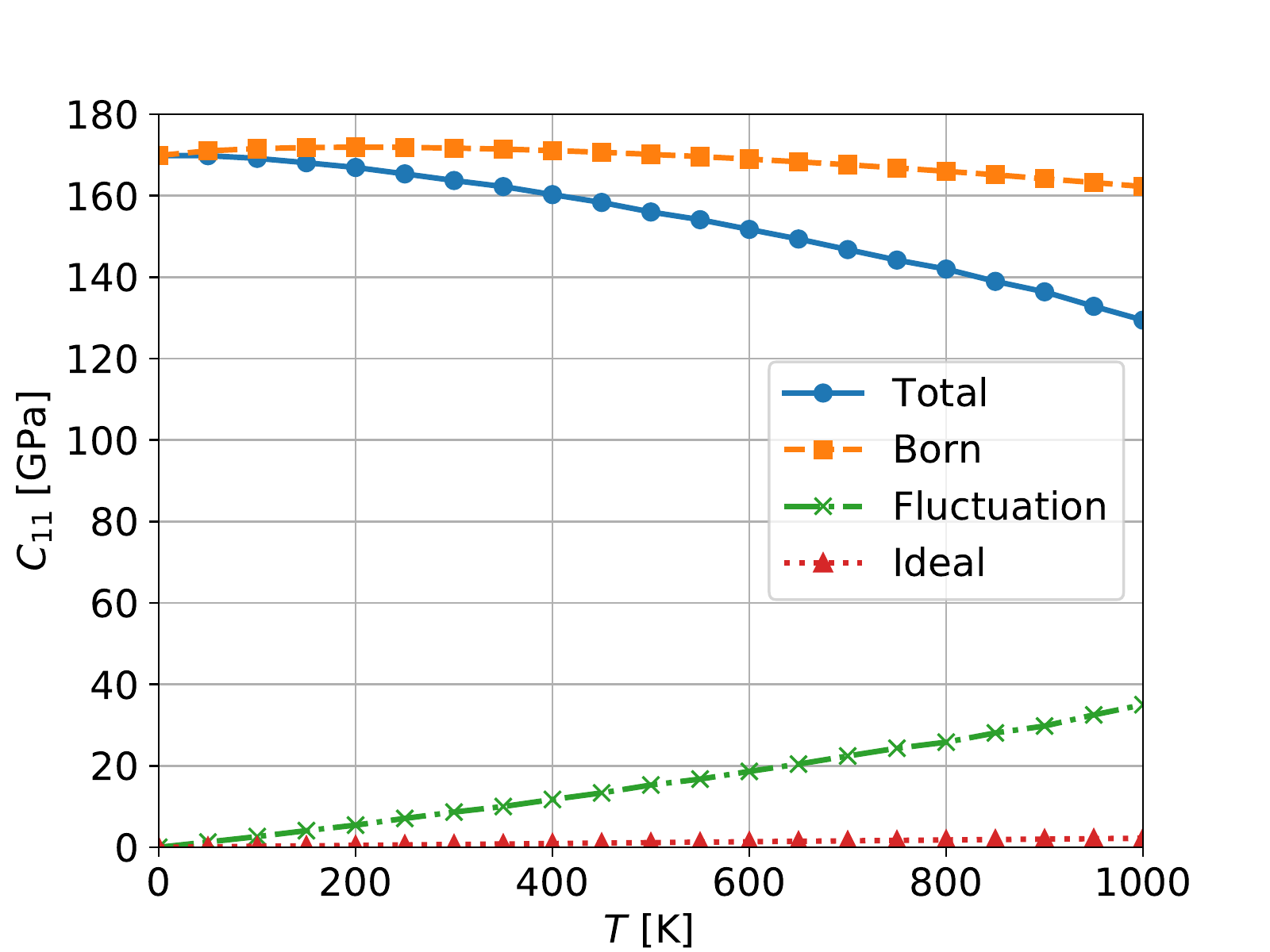}\includegraphics[scale=0.38]{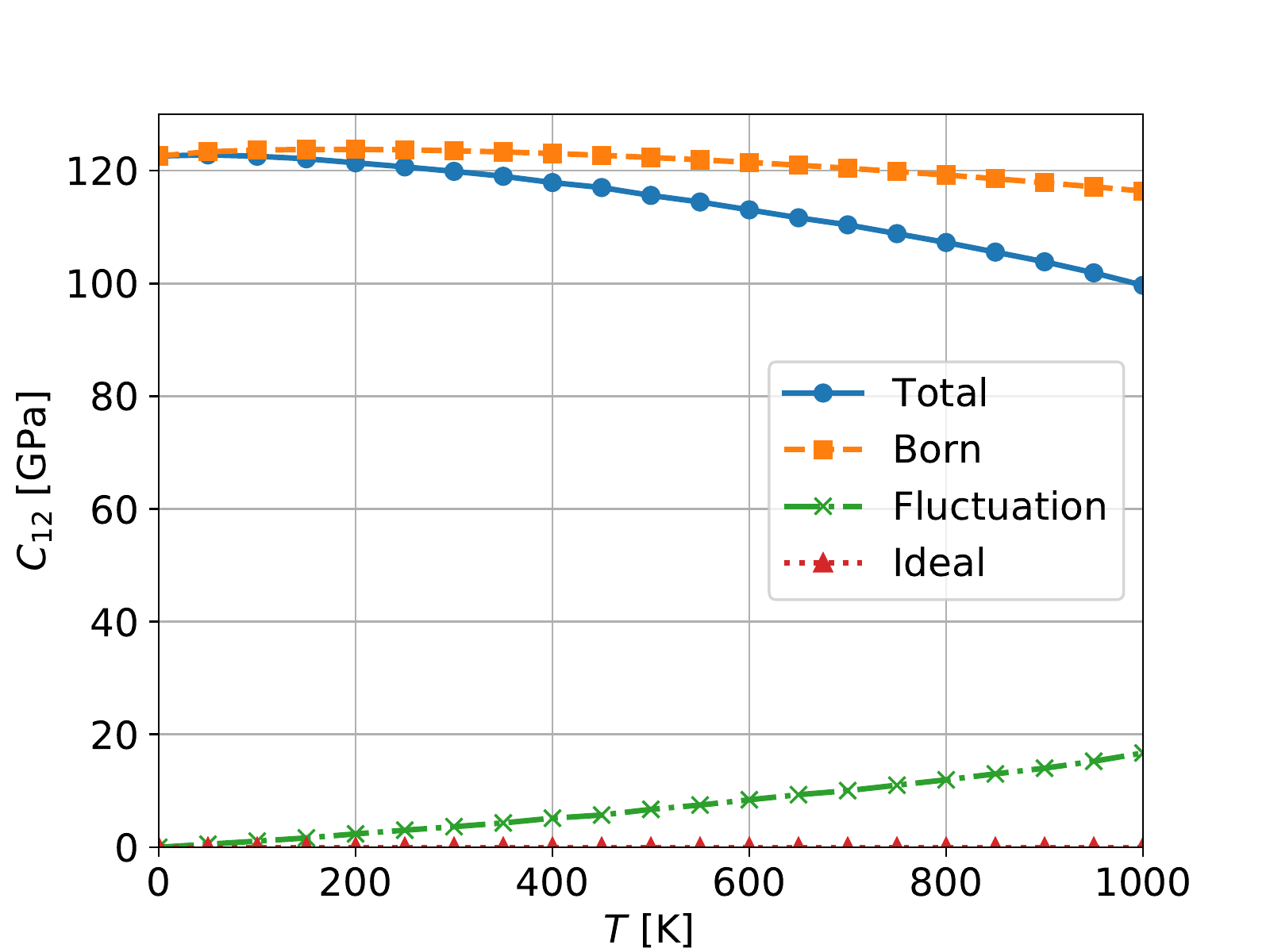}\includegraphics[scale=0.38]{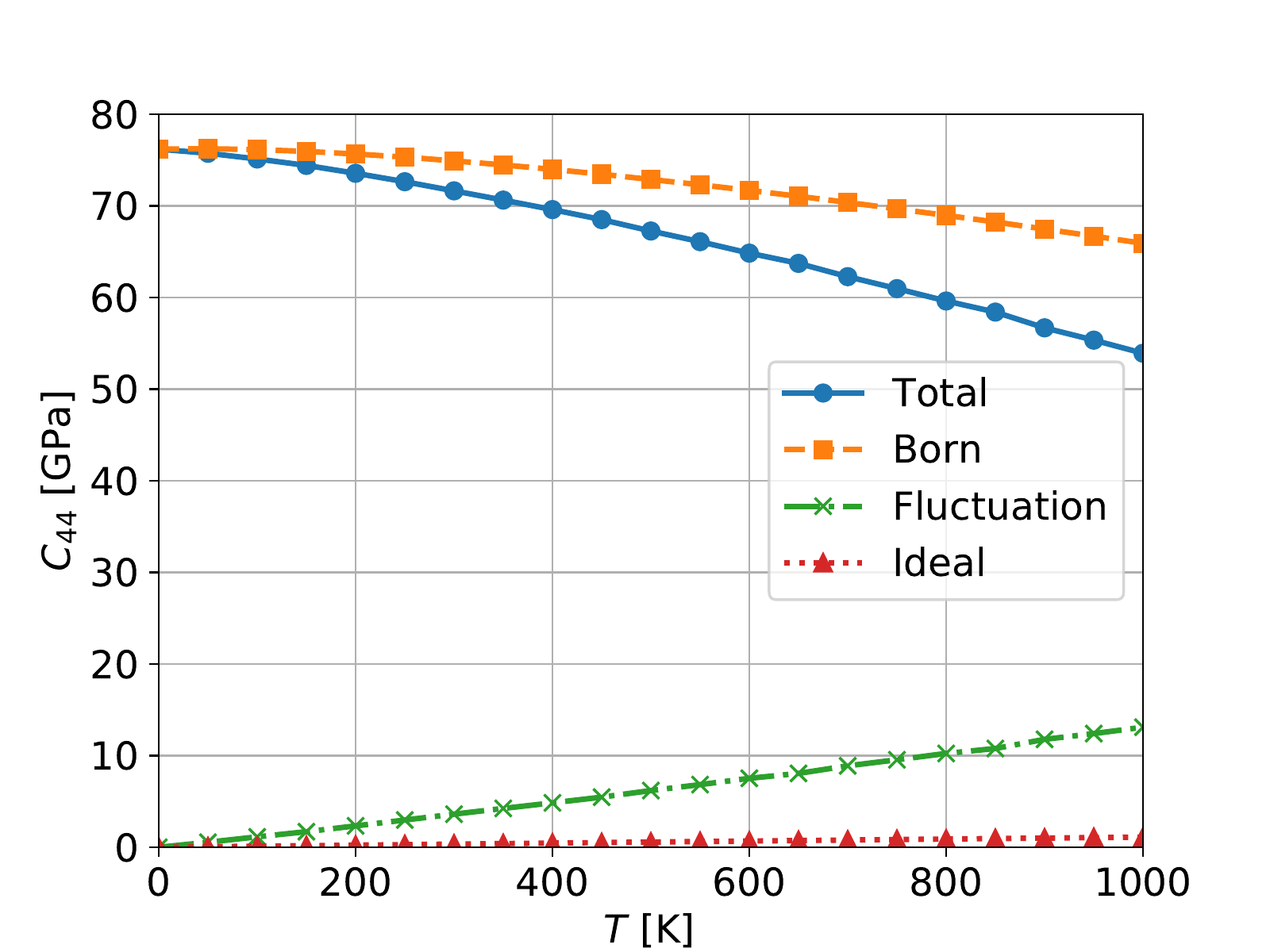}
\par\end{centering}
\caption{(Color online) The different terms contributing to the total isothermal elastic constants
$C_{11}$ (left pane), $C_{12}$ (middle pane) and $C_{44}$ (right
pane) of Copper as a function of temperature (solid blue line) - the
Born term (first term in eq. \ref{eq:elastic_micro}, dashed orange
line), Born stress fluctuation term (minus the second term in eq.
\ref{eq:elastic_micro}, dashed dotted green line) and the ideal gas
kinetic term (minus the last term in eq. \ref{eq:elastic_micro},
dotted red line). \label{fig:CT_contri}}
\end{figure*}
\begin{figure}
\begin{centering}
\includegraphics[scale=0.52]{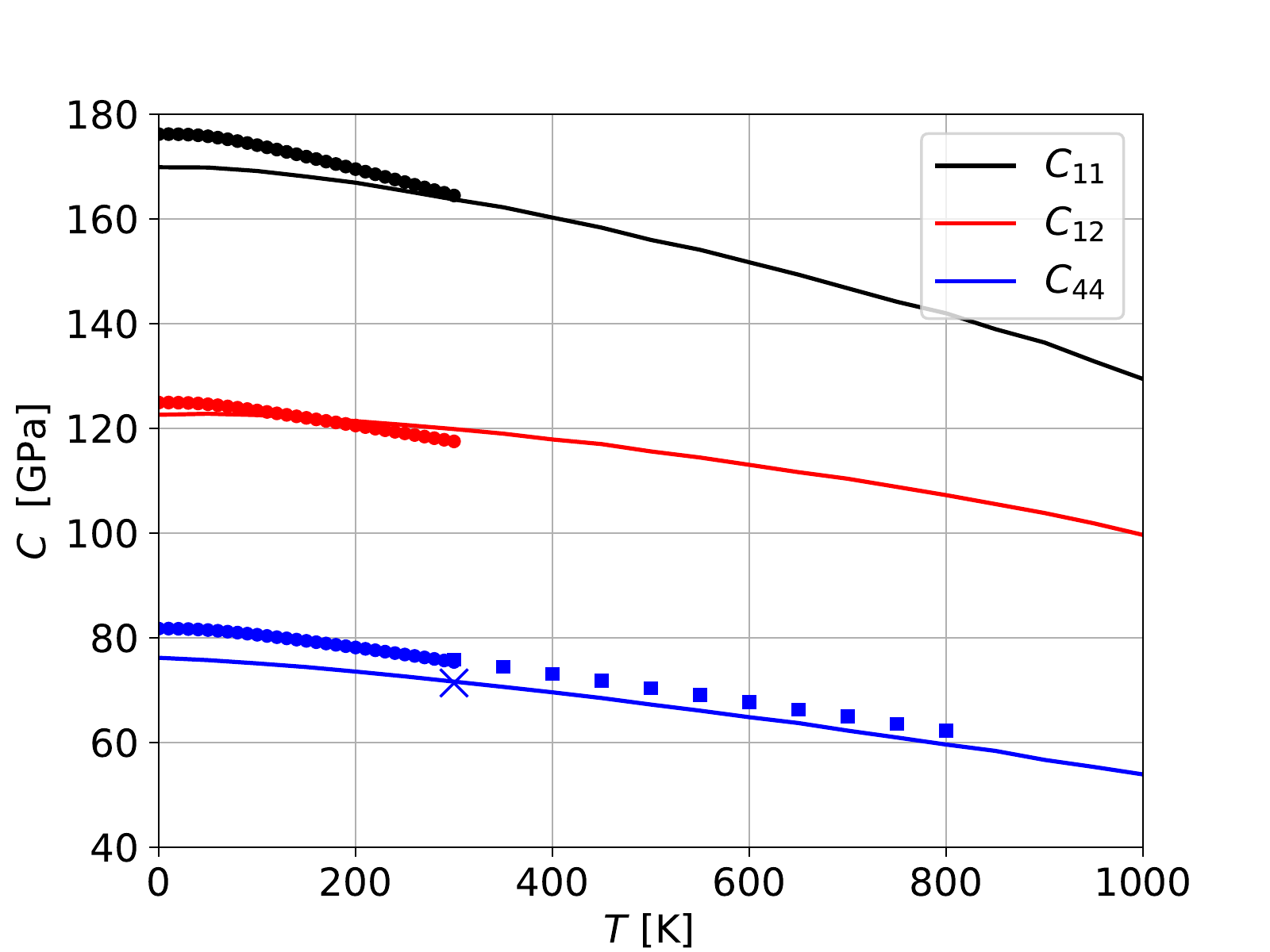}
\par\end{centering}
\caption{(Color online) The isothermal elastic constants $C_{11}$ (black), $C_{12}$ (red)
and $C_{44}$ (blue) of Copper as a function of temperature. Solid
lines represent molecular dynamics calculations performed in this
work (see text for details), circles represent experimental values
in the temperature range 0-300K given in Ref. \cite{overton1955temperature}
and squares represent experimental values in the range 300-800K given
in Ref. \cite{chang1966temperature} for $C_{44}$. The blue X data point represents a calculated $C_{44}$ result given in Ref.  \cite{rassoulinejad2016evaluation}.\label{fig:C_tot_expr}}
\end{figure}


Detailed molecular dynamics simulations are performed, in order to demonstrate and analyze the formalism developed in the previous sections, which we employ for the calculation of the elasticity tensor of copper.

The results presented in this work were obtained using a newly developed
molecular dynamics code. The computational model employs symplectic numerical integrators, which preserve phase space measures
of non-Hamiltonian dynamics \cite{martyna1994constant,martyna1996explicit,tuckerman2001non,tuckerman2006liouville,tuckerman2010statistical,allen2017computer},
incorporates Nose-Hoover thermostat chains in the NVT and NPT ensembles
\cite{martyna1992nose,nose1984unified,hoover1985canonical}, solved usin high order Suzuki-Yoshida decomposition \cite{yoshida1990construction,suzuki1991general}.
The equations of motions  are detailed in appendices \ref{app:nvt}
and \ref{app:npt}, for the NVT and NPT ensembles, respectively.

The isothermal elastic constants of Copper were calculated
as a function of temperature using a series
of molecular dynamics calculations in the NVT ensemble employing a widely used EAM potential by Mishin et al. \cite{mishin2001structural}.
This potential is calibrated so that the calculated zero temperature fcc lattice
constant is $L\left(T=0\right)=3.615\mathrm{\mathring{A}}$ and the
elastic constants are given by $C_{11}=169.9\text{GPa}$, $C_{12}=122.6\text{GPa}$
and $C_{44}=76.2\text{GPa}$ (the Voigt notation will be used throughout
this section). In the simulations presented below, a cubical system
with 500 atoms was used with their initial positions on an fcc lattice,
and initial velocities sampled from a Maxwell distribution at the
appropriate temperature. Periodic boundary conditions were used. The
integration of the equations of motion was performed with a time step
of $2\text{fs}$ for $2\times10^{5}$ steps. A Nose-Hoover thermostat
chain of size $M=10$ and a relaxation time of $\tau_{T}=50\text{fs}$
was applied.

The NVT calculations require prior evaluation of the equilibrium volume at zero stress at the various temperatures. These volumes can be evaluated using known thermal expansion coefficient, but we re-evaluated them using a series of zero stress NPT calculations. 
These yield the lattice constant as a function of temperature, $L\left(T\right)$.
The resulting equilibrium thermal expansion ratio, $L\left(T\right)/L\left(0\right)-1$,
is shown as a function of temperature in Fig. \ref{fig:L_T}. An analysis
of the thermalization and convergence of a selected NPT simulation
at $T=300$K is presented in detail in figures \ref{fig:npt_time}-\ref{fig:npt_p_tensor}.
It is evident that convergence is achieved relatively fast with $N=10^{4}$
MD steps. The instantaneous pressure tensor is shown in Fig. \ref{fig:npt_p_tensor},
demonstrating that the pressure is equilibrated to zero.
The NPT simulations were initialized with an fcc lattice with the
zero temperature lattice constant $L\left(T=0\right)=3.615\mathrm{\mathring{A}}$,
which is then subsequently relaxed to the equilibrium value $L\left(T\right)$
at zero pressure (as seen in Fig. \ref{fig:npt_time}). A barostat,
coupled to a Nose-Hoover thermostat chain of size $M'=10$ with a
relaxation time of $\tau_{P}=500\text{fs}$ was used. 

Results of a specific NVT simulation with $T=300$K and a volume chosen
such that the total pressure is zero, are shown in figures \ref{fig:NVT_born_stress_fluc}-\ref{fig:NVT_C_born}.
Fig. \ref{fig:NVT_born_stress_fluc} shows the components of the Born
stress tensor (defined in eq. \ref{eq:sigma_born_F}) throughout the
simulation. As expected, only the diagonal ideal gas terms are  nonzero (since the total pressure is zero,
from eq. \ref{eq:sigma_cauchy_id} we have $\left\langle \sigma_{\alpha\beta}^{B}\right\rangle =\delta_{\alpha\beta}k_{B}TN/V$).
Fig. \ref{fig:NVT_C_born} shows the Born elastic constants $C_{11}^{B},C_{12}^{B},C_{44}^{B}$
given in eq. \ref{eq:cborn_micro}, as in Fig. \ref{fig:npt_time}\@.
It is evident that the Born elastic constants are converged relatively
fast, at about $10^{4}$ MD steps. 

In Fig. \ref{fig:econs} we show the conservation of the energy like
quantities for both the NPT (given in eq. \ref{eq:econst_npt}) and
NVT calcualtions and NVT (given in eq. \ref{eq:econst_nvt}) throughout
the simulations discussed above. It is evident that a good conservation
of about $10^{-5}$ is maintained throughout the simulation, which
highlights the great advantages of measure preserving integrators
of the non-Hamltonian dynamical systems \cite{martyna1996explicit,tuckerman2001non}.

In Fig. \ref{fig:CT_contri} the isothermal elastic constants as a
function of temperature are plotted, together with the contributions
of the Born term, the Born stress fluctuation term and the ideal gas
kinetic term (see eq. \ref{eq:elastic_micro}). As expected, it is
evident that the reduction of the elastic constant with increasing
temperature is mainly due to the stress fluctuations term, which is
larger at higher temperatures. It is also evident that in the temperature
range studied here, the ideal gas contribution is negligible.

Finally, in Fig. \ref{fig:C_tot_expr}, we compare the calculated
isothermal elastic constants as a function of temperature, with experimental
results. In Ref. \cite{overton1955temperature}, experimental results
for isothermal elastic constants $C_{11},C_{12},C_{44}$ are given
in the temperature range 0-300K. In Ref. \cite{chang1966temperature},
experimental results for adiabatic elastic constants are given in
the range 300-800K. Since the isothermal and adiabatic elastic constants
are identical for $C_{44}$ , we can use these experimental results
directly. On the other hand, the isothermal and adiabatic elastic
constants $C_{11},C_{12}$ are not identical \cite{wang2010first,gulseren2002high,pham2011finite}\cite{wang2010first,gulseren2002high,pham2011finite},
and were not compared here. It is evident that the agreement between
the calculated and experimental values is relatively good, especially
given the fact that the Copper potential we used \cite{mishin2001structural}
was calibrated to slightly different values at $T=0$K (the experimental
values of Ref. \cite{overton1955temperature} are $C_{11}=176.2\text{GPa}$,
$C_{11}=124.94\text{GPa}$ and $C_{11}=81.77\text{GPa}$). The values
that were used to calibrate the tabulated potential (at $T=0)$, are
reproduced by our calculations to 5 significant digits. In addition, in Ref. \cite{rassoulinejad2016evaluation}  adiabatic elastic constants are calculated by molecular dynamics simulations (using LAMMPS MD code \cite{plimpton1995fast}) at $T=300$K, using the same EAM potential and employing the explicit deformation method. This result for $C_{44}$ is also shown in Fig. \ref{fig:C_tot_expr}, showing a very good agreement with our calculation (less than 0.5\%).

and at  results calculated with the same potential (and using LAMMPS \cite{plimpton1995fast}), The blue X data point represents the computational $C_{44}$ result given in Ref.  \cite{rassoulinejad2016evaluation}

\section{Summary}

The elasticity tensor of copper, modeled with a realistic tabulated EAM
potential, was calculated using a single molecular dynamics simulation
in the NVT ensamble, employing the stress-stress fluctuation formulation.
It was shown that such calculations are accurate and robust and converge within a few thousand MD steps, with a relatively small number
of particles (a few hundreds). The calculations were performed in
the temperature range of 0-1000K and compared to experimental values,
showing a good quantitative and qualitative agreement. The various
thermal contributions to the values of the elastic constants were
studied as a function of temperature. 

The results suggest that this method can be applied to calculate local elastic constants of real
crystalline materials for local regions embedded within large crystalline structure. In addition, it can improve
the calibration process of inter-atomic potentials, which typically employ
the explicit deformation method \cite{rassoulinejad2016evaluation,desai2019lammps,clavier2017computation,griebel2004molecular, quesnel1993elastic, manevitch2004elastic, vashishta2007interaction, pei2010mechanical}, that requires the generation of a consistent set of simulations
under varying deformations, in order to obtain the elasticity tensor. The use of a single simulation, other than being simpler, convenient and accurate, has a lower risk for the spontaneous appearance of defects and other phases during direct force calculations.

\bibliographystyle{unsrt}
\bibliography{datab}

\begin{thebibliography}{10}

\bibitem{daw1984embedded}
Murray~S Daw and Michael~I Baskes.
\newblock Embedded-atom method: Derivation and application to impurities,
  surfaces, and other defects in metals.
\newblock {\em Physical Review B}, 29(12):6443, 1984.

\bibitem{wolf1992temperature}
Ralph~J Wolf, Khalid~A Mansour, Myung~W Lee, and John~R Ray.
\newblock Temperature dependence of elastic constants of embedded-atom models
  of palladium.
\newblock {\em Physical Review B}, 46(13):8027, 1992.

\bibitem{ccaugin1999thermal}
T~{\c{C}}a{\u{g}}{\i}n, G~Dereli, M~Uludo{\u{g}}an, and MEHMET Tomak.
\newblock Thermal and mechanical properties of some fcc transition metals.
\newblock {\em Physical review B}, 59(5):3468, 1999.

\bibitem{chantasiriwan1996higher}
Somchart Chantasiriwan and Frederick Milstein.
\newblock Higher-order elasticity of cubic metals in the embedded-atom method.
\newblock {\em Physical Review B}, 53(21):14080, 1996.

\bibitem{ray1985molecular}
John~R Ray, Michael~C Moody, and Aneesur Rahman.
\newblock Molecular dynamics calculation of elastic constants for a crystalline
  system in equilibrium.
\newblock {\em Physical Review B}, 32(2):733, 1985.

\bibitem{squire1969isothermal}
DR~Squire, AC~Holt, and WG~Hoover.
\newblock Isothermal elastic constants for argon. theory and monte carlo
  calculations.
\newblock {\em Physica}, 42(3):388--397, 1969.

\bibitem{lutsko1989generalized}
JF~Lutsko.
\newblock Generalized expressions for the calculation of elastic constants by
  computer simulation.
\newblock {\em Journal of applied physics}, 65(8):2991--2997, 1989.

\bibitem{clavier2017computation}
Germain Clavier, Nicolas Desbiens, Emeric Bourasseau, V{\'e}ronique Lachet,
  Nad{\`e}ge Brusselle-Dupend, and Bernard Rousseau.
\newblock Computation of elastic constants of solids using molecular
  simulation: comparison of constant volume and constant pressure ensemble
  methods.
\newblock {\em Molecular Simulation}, 43(17):1413--1422, 2017.

\bibitem{lips2018stress}
Dominik Lips and Philipp Maass.
\newblock Stress-stress fluctuation formula for elastic constants in the npt
  ensemble.
\newblock {\em Physical Review E}, 97(5):053002, 2018.

\bibitem{rassoulinejad2016evaluation}
Seyed~Moein Rassoulinejad-Mousavi, Yijin Mao, and Yuwen Zhang.
\newblock Evaluation of copper, aluminum, and nickel interatomic potentials on
  predicting the elastic properties.
\newblock {\em Journal of Applied Physics}, 119(24):244304, 2016.

\bibitem{griebel2004molecular}
Michael Griebel and Jan Hamaekers.
\newblock Molecular dynamics simulations of the elastic moduli of
  polymer--carbon nanotube composites.
\newblock {\em Computer methods in applied mechanics and engineering},
  193(17-20):1773--1788, 2004.

\bibitem{quesnel1993elastic}
DJ~Quesnel, DS~Rimai, and LP~DeMejo.
\newblock Elastic compliances and stiffnesses of the fcc lennard-jones solid.
\newblock {\em Physical Review B}, 48(10):6795, 1993.

\bibitem{manevitch2004elastic}
Oleg~L Manevitch and Gregory~C Rutledge.
\newblock Elastic properties of a single lamella of montmorillonite by
  molecular dynamics simulation.
\newblock {\em The Journal of Physical Chemistry B}, 108(4):1428--1435, 2004.

\bibitem{vashishta2007interaction}
Priya Vashishta, Rajiv~K Kalia, Aiichiro Nakano, and Jos{\'e}~Pedro Rino.
\newblock Interaction potential for silicon carbide: A molecular dynamics study
  of elastic constants and vibrational density of states for crystalline and
  amorphous silicon carbide.
\newblock {\em Journal of applied physics}, 101(10):103515, 2007.

\bibitem{pei2010mechanical}
Qing-Xiang Pei, Yong-Wei Zhang, and Vivek~B Shenoy.
\newblock Mechanical properties of methyl functionalized graphene: a molecular
  dynamics study.
\newblock {\em Nanotechnology}, 21(11):115709, 2010.

\bibitem{desai2019lammps}
Saaketh Desai and Alejandro Strachan.
\newblock Lammps driver tool for potential calibration.
\newblock 2019.

\bibitem{huang2006thermoelastic}
L~Huang, Levente Vitos, SK~Kwon, B{\"o}rje Johansson, and Rajeev Ahuja.
\newblock Thermoelastic properties of random alloys from first-principles
  theory.
\newblock {\em Physical Review B}, 73(10):104203, 2006.

\bibitem{keuter2019qualitative}
Philipp Keuter, Denis Music, Volker Schnabel, Michael Stuer, and Jochen~M
  Schneider.
\newblock From qualitative to quantitative description of the anomalous
  thermoelastic behavior of v, nb, ta, pd and pt.
\newblock {\em Journal of Physics: Condensed Matter}, 31(22):225402, 2019.

\bibitem{parrinello1982strain}
M~Parrinello and A~Rahman.
\newblock Strain fluctuations and elastic constants.
\newblock {\em The Journal of Chemical Physics}, 76(5):2662--2666, 1982.

\bibitem{gusev1996fluctuation}
Andrei~A Gusev, Marcel~M Zehnder, and Ulrich~W Suter.
\newblock Fluctuation formula for elastic constants.
\newblock {\em Physical Review B}, 54(1):1, 1996.

\bibitem{shinoda2004rapid}
Wataru Shinoda, Motoyuki Shiga, and Masuhiro Mikami.
\newblock Rapid estimation of elastic constants by molecular dynamics
  simulation under constant stress.
\newblock {\em Physical Review B}, 69(13):134103, 2004.

\bibitem{martyna1994constant}
Glenn~J Martyna, Douglas~J Tobias, and Michael~L Klein.
\newblock Constant pressure molecular dynamics algorithms.
\newblock {\em The Journal of chemical physics}, 101(5):4177--4189, 1994.

\bibitem{martyna1996explicit}
Glenn~J Martyna, Mark~E Tuckerman, Douglas~J Tobias, and Michael~L Klein.
\newblock Explicit reversible integrators for extended systems dynamics.
\newblock {\em Molecular Physics}, 87(5):1117--1157, 1996.

\bibitem{tuckerman2001non}
Mark~E Tuckerman, Yi~Liu, Giovanni Ciccotti, and Glenn~J Martyna.
\newblock Non-hamiltonian molecular dynamics: Generalizing hamiltonian phase
  space principles to non-hamiltonian systems.
\newblock {\em The Journal of Chemical Physics}, 115(4):1678--1702, 2001.

\bibitem{tuckerman2006liouville}
Mark~E Tuckerman, Jos{\'e} Alejandre, Roberto L{\'o}pez-Rend{\'o}n, Andrea~L
  Jochim, and Glenn~J Martyna.
\newblock A liouville-operator derived measure-preserving integrator for
  molecular dynamics simulations in the isothermal--isobaric ensemble.
\newblock {\em Journal of Physics A: Mathematical and General}, 39(19):5629,
  2006.

\bibitem{tuckerman2010statistical}
Mark Tuckerman.
\newblock {\em Statistical mechanics: theory and molecular simulation}.
\newblock Oxford university press, 2010.

\bibitem{allen2017computer}
Michael~P Allen and Dominic~J Tildesley.
\newblock {\em Computer simulation of liquids}.
\newblock Oxford university press, 2017.

\bibitem{martyna1992nose}
Glenn~J Martyna, Michael~L Klein, and Mark Tuckerman.
\newblock Nos{\'e}--hoover chains: The canonical ensemble via continuous
  dynamics.
\newblock {\em The Journal of chemical physics}, 97(4):2635--2643, 1992.

\bibitem{nose1984unified}
Shuichi Nos{\'e}.
\newblock A unified formulation of the constant temperature molecular dynamics
  methods.
\newblock {\em The Journal of chemical physics}, 81(1):511--519, 1984.

\bibitem{hoover1985canonical}
William~G Hoover.
\newblock Canonical dynamics: Equilibrium phase-space distributions.
\newblock {\em Physical review A}, 31(3):1695, 1985.

\bibitem{mishin2001structural}
Yu~Mishin, MJ~Mehl, DA~Papaconstantopoulos, AF~Voter, and JD~Kress.
\newblock Structural stability and lattice defects in copper: Ab initio,
  tight-binding, and embedded-atom calculations.
\newblock {\em Physical Review B}, 63(22):224106, 2001.

\bibitem{thurston1964wave}
RN~Thurston.
\newblock Wave propagation in fluids and normal solids.
\newblock In {\em Physical acoustics}, pages 1--110. Elsevier, 1964.

\bibitem{wallace1967thermoelasticity}
Duane~C Wallace.
\newblock Thermoelasticity of stressed materials and comparison of various
  elastic constants.
\newblock {\em Physical Review}, 162(3):776, 1967.

\bibitem{brugger1964thermodynamic}
K~Brugger.
\newblock Thermodynamic definition of higher order elastic coefficients.
\newblock {\em Physical Review}, 133(6A):A1611, 1964.

\bibitem{wallace1972thermodynamics}
Duane~C. Wallace.
\newblock {\em Thermodynamics of crystals}.
\newblock Wiley New York, 1972.

\bibitem{ray1984statistical}
John~R Ray and Aneesur Rahman.
\newblock Statistical ensembles and molecular dynamics studies of anisotropic
  solids.
\newblock {\em The Journal of chemical physics}, 80(9):4423--4428, 1984.

\bibitem{wojciechowski1987constant}
KW~Wojciechowski.
\newblock Constant thermodynamic tension monte carlo studies of elastic
  properties of a two-dimensional system of hard cyclic hexamers.
\newblock {\em Molecular Physics}, 61(5):1247--1258, 1987.

\bibitem{ray1988elastic}
John~R Ray.
\newblock Elastic constants and statistical ensembles in molecular dynamics.
\newblock {\em Computer physics reports}, 8(3):109--151, 1988.

\bibitem{van2005isothermal}
Kevin Van~Workum, Kenji Yoshimoto, Juan~J de~Pablo, and Jack~F Douglas.
\newblock Isothermal stress and elasticity tensors for ions and point dipoles
  using ewald summations.
\newblock {\em Physical Review E}, 71(6):061102, 2005.

\bibitem{barrat2006microscopic}
J-L Barrat.
\newblock Microscopic elasticity of complex systems.
\newblock In {\em Computer Simulations in Condensed Matter Systems: From
  Materials to Chemical Biology Volume 2}, pages 287--307. Springer, 2006.

\bibitem{born1954dynamical}
Max Born and Kun Huang.
\newblock {\em Dynamical theory of crystal lattices}.
\newblock Clarendon press, 1954.

\bibitem{overton1955temperature}
WC~Overton~Jr and John Gaffney.
\newblock Temperature variation of the elastic constants of cubic elements. i.
  copper.
\newblock {\em Physical Review}, 98(4):969, 1955.

\bibitem{chang1966temperature}
YA~Chang and L~Himmel.
\newblock Temperature dependence of the elastic constants of cu, ag, and au
  above room temperature.
\newblock {\em Journal of Applied Physics}, 37(9):3567--3572, 1966.

\bibitem{yoshida1990construction}
Haruo Yoshida.
\newblock Construction of higher order symplectic integrators.
\newblock {\em Physics letters A}, 150(5-7):262--268, 1990.

\bibitem{suzuki1991general}
Masuo Suzuki.
\newblock General theory of fractal path integrals with applications to
  many-body theories and statistical physics.
\newblock {\em Journal of Mathematical Physics}, 32(2):400--407, 1991.

\bibitem{wang2010first}
Y~Wang, JJ~Wang, H~Zhang, VR~Manga, SL~Shang, LQ~Chen, and ZK~Liu.
\newblock A first-principles approach to finite temperature elastic constants.
\newblock {\em Journal of Physics: Condensed Matter}, 22(22):225404, 2010.

\bibitem{gulseren2002high}
O~G{\"u}lseren and RE~Cohen.
\newblock High-pressure thermoelasticity of body-centered-cubic tantalum.
\newblock {\em Physical Review B}, 65(6):064103, 2002.

\bibitem{pham2011finite}
Hieu~H Pham, Michael~E Williams, Patrick Mahaffey, Miladin Radovic, Raymundo
  Arroyave, and Tahir Cagin.
\newblock Finite-temperature elasticity of fcc al: Atomistic simulations and
  ultrasonic measurements.
\newblock {\em Physical Review B}, 84(6):064101, 2011.

\bibitem{plimpton1995fast}
Steve Plimpton.
\newblock Fast parallel algorithms for short-range molecular dynamics.
\newblock {\em Journal of computational physics}, 117(1):1--19, 1995.

\end{thebibliography}

\pagebreak{}

\appendix

\section{Elastic constants in the canonical ensemble\label{app:Elastic-constant}}

In the canonical ensemble, the free energy is given by:
\begin{equation}
\mathcal{F}=-k_{B}T\ln Z,
\end{equation}
where the canonical partition function is given by:
\begin{align}
Z & =\frac{1}{h^{3N}}\int d^{3}\boldsymbol{p}^{N}\int_{V}d^{3}\boldsymbol{r}^{N}e^{-\beta\mathcal{H}\left(\boldsymbol{r}^{N},\boldsymbol{p}^{N}\right)}\nonumber \\
 & =\frac{1}{\prod_{i}^{N}\Lambda_{i}^{3}}\int_{V}d^{3}\boldsymbol{r}^{N}e^{-\beta\mathcal{V}\left(\boldsymbol{r}^{N}\right)},
\end{align}
where $\beta=1/k_{B}T$ and $\Lambda_{i}=\left(\frac{2\pi\hbar^{2}\beta}{m_{i}}\right)^{\frac{1}{2}}$
is the thermal de-Broglie wavelength. We note that in order calculate
strain derivatives (i.e. as in equations \ref{eq:sigma_cauchy}-\ref{eq:elastic_def}),
the $\frac{1}{\prod_{i}^{N}\Lambda_{i}^{3}}$ factor can be dropped,
so that for the derivations presented in this appendix, we re-denote:
\begin{equation}
Z=\int_{V}d^{3}\boldsymbol{r}^{N}e^{-\beta\mathcal{V}\left(\boldsymbol{r}^{N}\right)}.
\end{equation}
 Consider the canonical average of a configurational operator $A=A\left(\boldsymbol{r}^{N}\right)$:
\begin{equation}
\left\langle A\right\rangle =\frac{1}{Z}\int_{V}d^{3}\boldsymbol{r}^{N}A\left(\boldsymbol{r}^{N}\right)e^{-\beta\mathcal{V}\left(\boldsymbol{r}^{N}\right)}.\label{eq:A_av}
\end{equation}
In order to evaluate the strain derivative $\frac{\partial\left\langle A\right\rangle }{\partial\eta_{\alpha\beta}}$,
we write the integral in terms of the reference configuration coordinates
$\boldsymbol{R}$, since the strain derivative cannot be brought inside
the integral, because the deformed volume $V$ depends on $\eta$.
Therefore, we write:
\begin{equation}
\int_{V}d^{3}\boldsymbol{r}^{N}A\left(\boldsymbol{r}^{N}\right)e^{-\beta\mathcal{V}\left(\boldsymbol{r}^{N}\right)}=\int_{V_{0}}d^{3}\boldsymbol{R}^{N}\left(\det\boldsymbol{J}\right)^{N}Ae^{-\beta\mathcal{V}},
\end{equation}
where the integrand in the RHS is understood to be evaluated at $\boldsymbol{r}\left(\boldsymbol{R}\right)$.
We can now write the strain derivative:
\begin{align}
 & \frac{\partial}{\partial\eta_{\alpha\beta}}\int_{V}d^{3}\boldsymbol{r}^{N}A\left(\boldsymbol{r}^{N}\right)e^{-\beta\mathcal{V}\left(\boldsymbol{r}^{N}\right)}=\nonumber \\
 & \int_{V_{0}}d^{3}\boldsymbol{R}^{N}\frac{\partial}{\partial\eta_{\alpha\beta}}\left[\left(\det\boldsymbol{J}\right)^{N}Ae^{-\beta\mathcal{V}}\right],\label{eq:deriv_A_temp}
\end{align}
the integrand reads:
\begin{align*}
 & \frac{\partial}{\partial\eta_{\alpha\beta}}\left[\left(\det\boldsymbol{J}\right)^{N}Ae^{-\beta\mathcal{V}}\right]=\\
 & e^{-\beta\mathcal{V}}\left(\det\boldsymbol{J}\right)^{N}\left[A\frac{N}{\det\left(\boldsymbol{J}\right)}\frac{\partial\det\left(\boldsymbol{J}\right)}{\partial\eta_{\alpha\beta}}+\frac{\partial A}{\partial\eta_{\alpha\beta}}-\beta A\frac{\partial\mathcal{V}}{\partial\eta_{\alpha\beta}}\right].
\end{align*}
Using the well known rule for the derivative of a determinant:
\begin{equation}
\frac{\partial\det\left(\boldsymbol{M}\right)}{\partial M_{\alpha\beta}}=\det\left(\boldsymbol{M}\right)M_{\alpha\beta}^{-T},
\end{equation}
and eq. \ref{eq:eta_J}, it is readily shown that:
\begin{equation}
\frac{1}{\det\left(\boldsymbol{J}\right)}\frac{\partial\det\left(\boldsymbol{J}\right)}{\partial\eta_{\alpha\beta}}=\frac{1}{\det\left(\boldsymbol{J}\right)}D_{\alpha\beta}\det\left(\boldsymbol{J}\right)=\left(2\boldsymbol{\eta}+\boldsymbol{I}\right)_{\alpha\beta}^{-T}.
\end{equation}
As a result, eq. \ref{eq:deriv_A_temp} reads:
\begin{align}
 & \frac{\partial}{\partial\eta_{\alpha\beta}}\left(\int_{V}d^{3}\boldsymbol{r}^{N}A\left(\boldsymbol{r}^{N}\right)e^{-\beta\mathcal{V}\left(\boldsymbol{r}^{N}\right)}\right)=\nonumber \\
 & \int_{V_{0}}d^{3}\boldsymbol{R}^{N}\left(\det\boldsymbol{J}\right)^{N}\left[A\left(2\boldsymbol{\eta}+\boldsymbol{I}\right)_{\alpha\beta}^{-T}+\frac{\partial A}{\partial\eta_{\alpha\beta}}-\beta A\frac{\partial\mathcal{V}}{\partial\eta_{\alpha\beta}}\right]e^{-\beta\mathcal{V}}\nonumber \\
 & =\int_{V}d^{3}\boldsymbol{r}^{N}\left[A\left(2\boldsymbol{\eta}+\boldsymbol{I}\right)_{\alpha\beta}^{-T}+\frac{\partial A}{\partial\eta_{\alpha\beta}}-\beta A\frac{\partial\mathcal{V}}{\partial\eta_{\alpha\beta}}\right]e^{-\beta\mathcal{V}}.\label{eq:A_deriv_temp}
\end{align}
Using this for the particular case $A\equiv1$, one finds:
\begin{align}
\frac{1}{Z}\frac{\partial Z}{\partial\eta_{\alpha\beta}} & =\left(2\boldsymbol{\eta}+\boldsymbol{I}\right)_{\alpha\beta}^{-T}-\beta\left\langle \frac{\partial\mathcal{V}}{\partial\eta_{\alpha\beta}}\right\rangle .\label{eq:z_deriv}
\end{align}
Differentiation of eq. \ref{eq:A_av} using the derivative product
rule and equations \ref{eq:A_deriv_temp}-\ref{eq:z_deriv}, results
in the general strain derivative rule:
\begin{equation}
\frac{\partial\left\langle A\right\rangle }{\partial\eta_{\alpha\beta}}=\left\langle \frac{\partial A}{\partial\eta_{\alpha\beta}}\right\rangle -\beta\left[\left\langle A\frac{\partial\mathcal{V}}{\partial\eta_{\alpha\beta}}\right\rangle -\left\langle A\right\rangle \left\langle \frac{\partial\mathcal{V}}{\partial\eta_{\alpha\beta}}\right\rangle \right].\label{eq:strain_deriv_A}
\end{equation}

For an EAM potential of the form \ref{eq:EAM}, we can write:

\begin{align*}
\frac{\partial\mathcal{V}}{\partial\eta_{\alpha\beta}} & =\sum_{i}\frac{\partial\mathcal{V}}{\partial\boldsymbol{r}_{i}}\cdot\frac{\partial\boldsymbol{r}_{i}}{\partial\eta_{\alpha\beta}}=-\sum_{i}F_{i,\gamma}\frac{\partial r_{i,\gamma}}{\partial\eta_{\alpha\beta}}\\
 & =-\sum_{i}\sum_{j\neq i}F_{ij}\frac{r_{ij,\gamma}}{r_{ij}}\frac{\partial r_{i,\gamma}}{\partial\eta_{\alpha\beta}}\\
 & =-\frac{1}{2}\sum_{i}\sum_{j\neq i}\left(F_{ij}\frac{r_{ij,\gamma}}{r_{ij}}\frac{\partial r_{i,\gamma}}{\partial\eta_{\alpha\beta}}+F_{ji}\frac{r_{ji,\gamma}}{r_{ji}}\frac{\partial r_{j,\gamma}}{\partial\eta_{\alpha\beta}}\right)\\
 & =-\frac{1}{2}\sum_{i}\sum_{j\neq i}F_{ij}\frac{r_{ij,\gamma}}{r_{ij}}\frac{\partial r_{ij,\gamma}}{\partial\eta_{\alpha\beta}}=-\frac{1}{2}\sum_{i<j}\frac{F_{ij}}{r_{ij}}\frac{\partial r_{ij}^{2}}{\partial\eta_{\alpha\beta}},
\end{align*}
where in the last term, the summation is over ordered pairs. From eq. \ref{eq:strain_length} it is evident that: 
\begin{equation}
\frac{\partial r^{2}}{\partial\eta_{\alpha\beta}}=2R_{\alpha}R_{\beta},
\end{equation}
and, more generally, for any radial function $f=f\left(r\right)$,
we have:
\begin{equation}
D_{\alpha\beta}f=\frac{\partial f}{\partial\eta_{\alpha\beta}}=\frac{\partial f}{\partial\eta_{\beta\alpha}}=\frac{df}{dr}\frac{R_{\alpha}R_{\beta}}{r}.\label{eq:dfr_deta}
\end{equation}
Hence, we can finally write:
\begin{equation}
\frac{\partial\mathcal{V}}{\partial\eta_{\alpha\beta}}=\frac{\partial\mathcal{V}}{\partial\eta_{\beta\alpha}}=D_{\alpha\beta}\mathcal{V}=-\sum_{i<j}F_{ij}\frac{R_{ij,\alpha}R_{ij,\beta}}{r_{ij}}.\label{eq:dv_deta}
\end{equation}
From eq. \ref{eq:z_deriv} it follows that:

\begin{align}
\frac{\partial\mathcal{F}}{\partial\eta_{\alpha\beta}}= & \left\langle \frac{\partial\mathcal{V}}{\partial\eta_{\alpha\beta}}\right\rangle -k_{B}TN\left(2\boldsymbol{\eta}+\boldsymbol{I}\right)_{\alpha\beta}^{-T},\label{eq:dF_deta}
\end{align}
which is a symmetric tensor. Hence, eq. \ref{eq:dF_deta}, when evaluated
at zero strain, proves equations \ref{eq:sigma_cauchy_id} and \ref{eq:sigma_born_F}.

Next, in order to evaluate the isothermal elastic constant defined
by eq. \ref{eq:elastic_def}, consider the second derivative tensor:

\begin{equation}
D_{\alpha\beta}D_{\gamma\delta}\mathcal{F}=D_{\alpha\beta}\left[\left\langle \frac{\partial\mathcal{V}}{\partial\eta_{\gamma\delta}}\right\rangle -k_{B}TN\left(2\boldsymbol{\eta}+\boldsymbol{I}\right)_{\gamma\delta}^{-T}\right].\label{eq:df_deta2}
\end{equation}
In order to evaluate the second term, we use the following identity
for the derivative of a matrix inverse:

\begin{equation}
\frac{\partial M_{\delta\gamma}^{-1}}{\partial M_{\alpha\beta}}=-M_{\delta\alpha}^{-1}M_{\beta\gamma}^{-1},
\end{equation}
which results in the relation:

\begin{align}
D_{\alpha\beta}\left[\left(2\boldsymbol{\eta}+\boldsymbol{I}\right)_{\gamma\delta}^{-T}\right] & =-\left(2\boldsymbol{\eta}+\boldsymbol{I}\right)_{\delta\alpha}^{-1}\left(2\boldsymbol{\eta}+\boldsymbol{I}\right)_{\beta\gamma}^{-1}\nonumber \\
 & -\left(2\boldsymbol{\eta}+\boldsymbol{I}\right)_{\delta\beta}^{-1}\left(2\boldsymbol{\eta}+\boldsymbol{I}\right)_{\alpha\gamma}^{-1}.\label{eq:kinetic_term_deta}
\end{align}
In order to evaluate the first term in \ref{eq:df_deta2}, we use
the identity \ref{eq:strain_deriv_A}:

\begin{align}
 & \frac{\partial}{\partial\eta_{\alpha\beta}}\left\langle \frac{\partial\mathcal{V}}{\partial\eta_{\gamma\delta}}\right\rangle =\nonumber \\
 & \left\langle \frac{\partial^{2}\mathcal{V}}{\partial\eta_{\alpha\beta}\partial\eta_{\gamma\delta}}\right\rangle -\beta\left[\left\langle \frac{\partial\mathcal{V}}{\partial\eta_{\gamma\delta}}\frac{\partial\mathcal{V}}{\partial\eta_{\alpha\beta}}\right\rangle -\left\langle \frac{\partial\mathcal{V}}{\partial\eta_{\gamma\delta}}\right\rangle \left\langle \frac{\partial\mathcal{V}}{\partial\eta_{\alpha\beta}}\right\rangle \right]\nonumber \\
 & =\left\langle \frac{\partial^{2}\mathcal{V}}{\partial\eta_{\alpha\beta}\partial\eta_{\gamma\delta}}\right\rangle -\frac{V^{2}}{k_{B}T}\left[\left\langle \sigma_{\alpha\beta}^{B}\sigma_{\gamma\delta}^{B}\right\rangle -\left\langle \sigma_{\alpha\beta}^{B}\right\rangle \left\langle \sigma_{\gamma\delta}^{B}\right\rangle \right].\label{eq:d2v_deta_fluc}
\end{align}
We note that the second term is a symmetric tensor with respect to
$\alpha\leftrightarrow\beta$ and $\gamma\leftrightarrow\delta$.
In order to evaluate the first term, we differentiate eq. \ref{eq:dv_deta}:
\begin{align}
 & \frac{\partial^{2}\mathcal{V}}{\partial\eta_{\alpha\beta}\partial\eta_{\gamma\delta}}=\frac{\partial}{\partial\eta_{\alpha\beta}}\left(-\sum_{i<j}F_{ij}\frac{R_{ij,\gamma}R_{ij,\delta}}{r_{ij}}\right)\nonumber \\
 & =-\sum_{i<j}\left[\frac{R_{ij,\gamma}R_{ij,\delta}}{r_{ij}}\frac{\partial F_{ij}}{\partial\eta_{\alpha\beta}}+R_{ij,\gamma}R_{ij,\delta}F_{ij}\frac{\partial}{\partial\eta_{\alpha\beta}}\left(\frac{1}{r_{ij}}\right)\right].\label{eq:d2F_deta}
\end{align}
Using the identity \ref{eq:dfr_deta}, we get:

\begin{align}
\frac{\partial}{\partial\eta_{\alpha\beta}}\left(\frac{1}{r}\right) & =-\frac{R_{\alpha}R_{\beta}}{r^{3}},\label{eq:drinv_deta}
\end{align}
and using eq. \ref{eq:Fij_eam} and the identity \ref{eq:dfr_deta}
gives:
\begin{align}
\frac{\partial F_{ij}}{\partial\eta_{\alpha\beta}} & =\left(v''\left(r_{ij}\right)+\left[F'\left(\rho_{i}\right)+F'\left(\rho_{j}\right)\right]\rho''\left(r_{ij}\right)\right)\frac{R_{ij,\alpha}R_{ij,\beta}}{r_{ij}}\nonumber \\
 & +\rho'\left(r_{ij}\right)\frac{\partial}{\partial\eta_{\alpha\beta}}\left[F'\left(\rho_{i}\right)+F'\left(\rho_{j}\right)\right].\label{eq:dFij_deta}
\end{align}
In order to evaluate the last term in eq. \ref{eq:dFij_deta}, we
write:
\begin{align}
\frac{\partial}{\partial\eta_{\alpha\beta}}F'\left(\rho_{i}\right) & =F''\left(\rho_{i}\right)\frac{\partial\rho_{i}}{\partial\eta_{\alpha\beta}}=F''\left(\rho_{i}\right)\sum_{k\neq i}\frac{\partial\rho}{\partial\eta_{\alpha\beta}}\bigg|_{r_{ik}}\nonumber \\
 & =F''\left(\rho_{i}\right)\sum_{k\neq i}\rho'\left(r_{ik}\right)\frac{R_{ik,\alpha}R_{ik,\beta}}{r_{ik}^{2}},\label{eq:dftag_deta}
\end{align}
where in the last step the identity \ref{eq:dfr_deta} was used again.
Plugging equations \ref{eq:drinv_deta}-\ref{eq:dFij_deta} back in
eq. \ref{eq:d2F_deta} gives:
\begin{align}
 & \frac{\partial^{2}\mathcal{V}}{\partial\eta_{\alpha\beta}\partial\eta_{\gamma\delta}}=D_{\alpha\beta}D_{\gamma\delta}\mathcal{V}=\sum_{i<j}X_{ij}\frac{R_{ij,\alpha}R_{ij,\beta}R_{ij,\gamma}R_{ij,\delta}}{r_{ij}^{2}}\nonumber \\
 & +\sum_{i}\left(\sum_{k\neq i}\frac{R_{ik,\alpha}R_{ik,\beta}}{r_{ik}^{2}}\rho'\left(r_{ik}\right)\right)\left(\sum_{k\neq i}\frac{R_{ik,\alpha}R_{ik,\beta}}{r_{ik}^{2}}\rho'\left(r_{ik}\right)\right),\label{eq:d2V_deta_final}
\end{align}
where $X_{ij}$ is defined by eq. \ref{eq:Xij}, and we have used
the fact that the resulting expression defines a symmetric tensor
with respect to $\alpha\leftrightarrow\beta$ and $\gamma\leftrightarrow\delta$.

Finally, when evaluated at zero strain ($\boldsymbol{\eta}=\boldsymbol{0}$,
$\boldsymbol{R}=\boldsymbol{r}$), the combination of equations \ref{eq:df_deta2},
\ref{eq:kinetic_term_deta}, \ref{eq:d2v_deta_fluc} and \ref{eq:d2V_deta_final}
proves the relations \ref{eq:elastic_micro} and \ref{eq:cborn_micro}-\ref{eq:g_def}.

\section{Equations of motion in the NVT ensemble\label{app:nvt}}

In this appendix we list the equations of motion of the widely used
Nose-Hoover-Chains method \cite{martyna1992nose,tuckerman2010statistical,allen2017computer},
which was used in the calculations performed in this work. We use
a chain of $M$ thermostats with coordinates $\eta_{j}$ and momenta
$p_{\eta_{j}}$. The equations of motions for the $i=1...N$ particles
are:

\begin{equation}
\frac{d\boldsymbol{r}_{i}}{dt}=\frac{\boldsymbol{p}_{i}}{m_{i}},\label{eq:app_eom_first}
\end{equation}
\begin{equation}
\frac{d\boldsymbol{p}_{i}}{dt}=\boldsymbol{F}_{i}-\left(\frac{p_{\eta_{1}}}{Q_{1}}\right)\boldsymbol{p}_{i}.
\end{equation}
The equations of motion for the $j=1...M$ thermostat variables are
given by:
\begin{equation}
\frac{d\eta_{j}}{dt}=\frac{p_{\eta_{j}}}{Q_{j}},
\end{equation}

\begin{equation}
\frac{dp_{\eta_{j}}}{dt}=\begin{cases}
G_{j}-\left(\frac{p_{\eta_{j+1}}}{Q_{j+1}}\right)p_{\eta_{j}}, & j=1...M-1\\
G_{M}, & j=M
\end{cases}\label{eq:eq:app_eom_last}
\end{equation}
where the forces are:
\begin{equation}
G_{j}=\begin{cases}
\sum_{i}\frac{p_{i}^{2}}{m_{i}}-gk_{B}T, & j=1\\
\frac{p_{\eta_{j-1}}^{2}}{Q_{j-1}}-k_{B}T, & j=2...M
\end{cases}
\end{equation}
with the number of degrees of freedom:
\begin{equation}
g=3\left(N-1\right).
\end{equation}
The thermostat ``masses'' $Q_{j}$ can be written in terms of a
thermostat relaxation timescale $\tau_{T}$ \cite{martyna1992nose}:
\begin{equation}
Q_{j}=\begin{cases}
gk_{B}T\tau_{T}^{2}, & j=1\\
k_{B}T\tau_{T}^{2}, & j=2...M
\end{cases}\label{eq:thermostat_masses}
\end{equation}
The non-Hamiltonian system \ref{eq:app_eom_first}-\ref{eq:eq:app_eom_last}
has the conserved quantity \cite{martyna1992nose,tuckerman2010statistical}:

\begin{equation}
E_{\text{cons}}=\mathcal{H}+\sum_{j=1}^{M}\frac{p_{\eta_{j}}^{2}}{2Q_{j}}+gk_{B}T\eta_{1}+\sum_{j=2}^{M}k_{B}T\eta_{j},\label{eq:econst_nvt}
\end{equation}
where $\mathcal{H}$ is the true Hamiltonian given by eq. \ref{eq:hamiltonian}.

\section{Equations of motion in the NPT ensemble\label{app:npt}}

In this appendix we list the equations of motion in the widely used
Nose-Hoover-Chains method in the isothermal-isobaric (NPT) ensemble
\cite{martyna1994constant,tuckerman2006liouville,tuckerman2010statistical,allen2017computer},
which was used in the calculations performed in this work. As in appendix
\ref{app:nvt}, we use a chain of $M$ thermostats whose coordinates
and momenta are $\eta_{j}$ and $p_{\eta_{j}}$, which are coupled
the the equations of motion of the $N$ particles. In the NPT ensemble,
the volume $V=V\left(t\right)$ is treated as a dynamical variable
via the generalized coordinate:
\begin{equation}
\epsilon\left(t\right)=\frac{1}{3}\ln\left(\frac{V\left(t\right)}{V\left(0\right)}\right),
\end{equation}
whose momentum $p_{\epsilon}$ is coupled to an additional chain of
$M'$ thermostats with coordinates $\eta_{j}'$ and momenta $p_{\eta_{j}}'$.
The equations of motions for the $i=1...N$ particles are:

\begin{equation}
\frac{d\boldsymbol{r}_{i}}{dt}=\frac{\boldsymbol{p}_{i}}{m_{i}}+\frac{p_{\epsilon}}{W}\boldsymbol{r}_{i},\label{eq:eq:app_eom_npt_first}
\end{equation}
\begin{equation}
\frac{d\boldsymbol{p}_{i}}{dt}=\boldsymbol{F}_{i}-\left(1+\frac{3}{g}\right)\left(\frac{p_{\epsilon}}{W}\right)\boldsymbol{p}_{i}-\left(\frac{p_{\eta_{1}}}{Q_{1}}\right)\boldsymbol{p}_{i},
\end{equation}
The equations of motion for the barostat variables are:

\begin{equation}
\frac{d\epsilon}{dt}=\frac{p_{\epsilon}}{W},
\end{equation}

\begin{equation}
\frac{dp_{\epsilon}}{dt}=3\left(P_{inst}V-PV\right)+\frac{3}{g}\sum_{i=1}^{N}\frac{p_{i}^{2}}{m_{i}}-\left(\frac{p_{\eta_{1}'}}{Q_{1}'}\right)p_{\epsilon},
\end{equation}
where $P$ is the applied external pressure and the instantaneous
pressure is given by the virial theorem:

\begin{equation}
P_{inst}V=\frac{1}{3}\sum_{i}\frac{p_{i}^{2}}{m_{i}}+\frac{1}{3}\sum_{i<j}F_{ij}r_{ij}.
\end{equation}
The equations of motion for the $j=1...M$ particle-coupled thermostat
variables are given by:
\begin{equation}
\frac{d\eta_{j}}{dt}=\frac{p_{\eta_{j}}}{Q_{j}},
\end{equation}

\begin{equation}
\frac{dp_{\eta_{j}}}{dt}=\begin{cases}
G_{j}-\left(\frac{p_{\eta_{j+1}}}{Q_{j+1}}\right)p_{\eta_{j}}, & j=1...M-1\\
G_{M}, & j=M
\end{cases}
\end{equation}
where the forces are:
\begin{equation}
G_{j}=\begin{cases}
\sum_{i}\frac{p_{i}^{2}}{m_{i}}-gk_{B}T, & j=1\\
\frac{p_{\eta_{j-1}}^{2}}{Q_{j-1}}-k_{B}T, & j=2...M
\end{cases}
\end{equation}
The equations of motion for the $j=1...M'$ barostat-coupled thermostat
variables are given by:

\begin{equation}
\frac{d\eta_{j}'}{dt}=\frac{p_{\eta_{j}'}}{Q_{j}'},
\end{equation}

\begin{equation}
\frac{dp_{\eta_{j}'}}{dt}=\begin{cases}
G_{j}'-\left(\frac{p_{\eta_{j+1}'}}{Q_{j+1}'}\right)p_{\eta_{j}'}, & j=1...M'-1\\
G_{M'}', & j=M'
\end{cases}\label{eq:eq:app_eom_npt_last}
\end{equation}
where the forces are:
\begin{equation}
G_{j}'=\begin{cases}
\frac{p_{\epsilon}^{2}}{W}-k_{B}T, & j=1\\
\frac{p_{\eta_{j-1}'}^{2}}{Q_{j-1}'}-k_{B}T, & j=2...M'
\end{cases}
\end{equation}
The particle-coupled thermostat masses $Q_{j}$ are given in terms
of a thermostat relaxation timescale $\tau_{T}$ as in \ref{eq:thermostat_masses},
while the barostat mass $W$ and the barostat-coupled thermostat masses
$Q_{j}'$ are written in terms of a barostat relaxation timescale
$\tau_{p}$ as  \cite{martyna1992nose, martyna1996explicit}:
\begin{equation}
W=gk_{B}T\tau_{p}^{2},
\end{equation}
\begin{equation}
Q_{j}'=k_{B}T\tau_{p}^{2}.
\end{equation}
The non-Hamiltonian system \ref{eq:eq:app_eom_npt_first}-\ref{eq:eq:app_eom_npt_last}
has the conserved quantity \cite{martyna1994constant,tuckerman2006liouville,tuckerman2010statistical}:

\begin{align}
E_{\text{cons}} & =\mathcal{H}+PV+\frac{p_{\epsilon}^{2}}{2W}+\sum_{j=1}^{M}\frac{p_{\eta_{j}}^{2}}{2Q_{j}}+\sum_{j=1}^{M'}\frac{p_{\eta_{j}'}^{2}}{2Q_{j}'}\nonumber \\
 & +gk_{B}T\eta_{1}+\sum_{j=2}^{M}k_{B}T\eta_{j}+\sum_{j=1}^{M'}k_{B}T\eta_{j}'.\label{eq:econst_npt}
\end{align}

\end{document}